\documentclass[a4paper,11pt]{article}
\pdfoutput=1 

\usepackage{jcappub} 

\usepackage[T1]{fontenc} 
\usepackage{units}
\usepackage{lineno}
\usepackage{endnotes}

\title{\boldmath Modulations of the Cosmic Muon Signal in Ten Years of Borexino Data}

\author{The Borexino Collaboration\hspace{15em}}

\author[a]{M.~Agostini}
\author[a]{K.~Altenm\"{u}ller}
\author[a]{S.~Appel}
\author[b]{V.~Atroshchenko}
\author[c]{Z.~Bagdasarian}
\author[d]{D.~Basilico}
\author[d]{G.~Bellini}
\author[e]{J.~Benziger}
\author[f]{D.~Bick}
% ...
\affiliation[a]{Physik-Department and Excellence Cluster Universe, Technische Universit\"at  M\"unchen, 85748 Garching, Germany}
\affiliation[b]{National Research Centre Kurchatov Institute, 123182 Moscow, Russia}
\affiliation[c]{Institut f\"ur Kernphysik, Forschungszentrum J\"ulich, 52425 J\"ulich, Germany}
\affiliation[d]{Dipartimento di Fisica, Universit\`a degli Studi e INFN, 20133 Milano, Italy}
\affiliation[e]{Chemical Engineering Department, Princeton University, Princeton, NJ 08544, USA}
\affiliation[f]{Institut f\"ur Experimentalphysik, Universit\"at Hamburg, 22761 Hamburg, Germany}
% ...

\abstract{We have measured the flux of cosmic muons in the Laboratori Nazionali del Gran Sasso at $\unit[3800]{m\,w.e.}$ to be $\unit[(3.432 \pm 0.003)\cdot 10^{-4}]{m^{-2}s^{-1}}$ based on ten years of Borexino data acquired between May 2007 and May 2017. A seasonal modulation with a period of $\unit[(366.3 \pm 0.6)]{d}$ and a relative amplitude of $(1.36 \pm0.04)\%$ is observed. The phase is measured to be $\unit[(181.7 \pm 0.4)]{d}$, corresponding to a maximum at the 1$^\mathrm{st}$ of July. Using data inferred from global atmospheric models, we show the muon flux to be positively correlated with the atmospheric temperature and measure the effective temperature coefficient $\alpha_\mathrm{T} = 0.90 \pm 0.02$. The origin of cosmic muons from pion and kaon decays in the atmosphere allows to interpret the effective temperature coefficient as an indirect measurement of the atmospheric kaon-to-pion production ratio $r_{\mathrm{K}/\pi} = 0.11^{+0.11}_{-0.07}$ for primary energies above $\unit[18]{TeV}$. We find evidence for a long-term modulation of the muon flux with a period of $\sim \unit[3000]{d}$ and a maximum in June 2012 that is not present in the atmospheric temperature data. A possible correlation between this modulation and the solar activity is investigated. The cosmogenic neutron production rate is found to show a seasonal modulation in phase with the cosmic muon flux but with an increased amplitude of $(2.6 \pm 0.4)\%$.}
\begin{document}
\maketitle
\flushbottom

\section{Introduction}
\label{sec:intro}
Cosmic muons are produced mainly in the decays of kaons and pions that originate from the interaction of primary cosmic rays with nuclei in the upper atmosphere~\cite{Gaisser}. For detectors situated deep underground, the flux of cosmic muons is strongly reduced. Only muons surpassing a certain threshold energy $E_\mathrm{thr}$ contribute, while lower energy muons are absorbed in the rock overburden. At great depths, the residual high energy muons must have been produced by parent mesons that decay in flight without any inelastic interactions and without elastic interactions of large momentum transfer before the decay. As a consequence, the density and temperature variations of the upper atmosphere that alter the mean free path of the decaying mesons introduce, in first approximation, a seasonal modulation of the underground muon flux, which has been investigated for many decades~\cite{Barrett}. Several experiments located at the Laboratori Nazionali del Gran Sasso (LNGS) in Italy such as MACRO~\cite{MACRO}, LVD~\cite{LVD,LVDII}, Borexino~\cite{BxFlux}, and GERDA~\cite{GERDA} and at other experimental sites, e.g. IceCube~\cite{IceCube}, MINOS~\cite{MINOS}, Double Chooz~\cite{DoubleChooz}, or Daya Bay~\cite{DAYA}, have studied this phenomenon. Compared to the previously published investigation based on four years of Borexino data acquired between 2007 and 2011~\cite{BxFlux}, the present analysis of ten years of data from 2007 to 2017 achieves a significantly better precision on the muon flux, on the modulation parameters, and on the effective temperature coefficient. In addition, we expand the former analysis by measuring the atmospheric kaon-to-pion production ratio, observe a long-term modulation of the cosmic muon flux, investigate a possible correlation between this modulation and the solar activity, and measure the seasonal modulation of the cosmogenic neutron production rate.\\
Borexino is an organic liquid scintillator detector situated at the LNGS, covered by a limestone overburden of $3800$\,m\,w.e.~\cite{BxDetector}. It is designed for the spectroscopy of low energy solar neutrinos that are detected via elastic scattering off electrons. Based on the data acquired after the start of data taking in May 2007, Borexino accomplished measurements of the solar $^7\mathrm{Be}$~\cite{7Be1,7Be2,pp7bepep,7Be3}, $^8\mathrm{B}$~\cite{8b1,8b2}, pep~\cite{pp7bepep,pep}, and pp neutrino fluxes~\cite{pp,pp7bepep}. 
The complete spectroscopy of neutrinos from the pp-chain performed with Borexino is now available in~\cite{nature2}. In addition, a limit on the flux of solar neutrinos produced in the CNO cycle~\cite{pp7bepep,pep} and a spectroscopic measurement of antineutrinos produced in radioactive decays within the Earth, the so-called geo-neutrinos~\cite{geo_1,geo_2,geo_3}, were performed.
Investigating the background to the neutrino analyses, Borexino further performed  detailed studies of high energy cosmic muons as well as of cosmogenic neutrons and radioactive isotopes from muon spallation on the detector materials~\cite{cosmogenics}.\\
The Borexino detector geometry allows to identify muons passing through a spherical volume with a cross section of 146\,$\mathrm{m}^2$. The detection efficiency is virtually independent of the muon's incident angle, resulting in minimum systematics when measuring the muon flux and its variations. Detailed air temperature data are provided by weather forecasting centers~\cite{ecmwf} for the location of the laboratory and can be used to investigate the correlation between the flux of high energy cosmic muons and the atmospheric temperature to determine the atmospheric temperature coefficient. \\
In this article, we present an analysis of the cosmic muon flux as measured by Borexino based on ten years of data. In section \ref{sec:Borexino}, we briefly introduce the Borexino detector. In section \ref{sec:Seas}, we report on the measured flux of cosmic muons and its seasonal modulation. In section \ref{sec:AtmoMod}, we introduce a model describing the expected relation between the flux of cosmic muons and the atmospheric temperature. In section \ref{sec:TempMod}, we present the modulation of the atmospheric temperature. In section \ref{sec:Corr}, we analyze the correlation between the flux of cosmic muons and the atmospheric temperature. In section \ref{sec:RKPI}, we use the inferred effective temperature coefficient to measure the kaon-to-pion production ratio in the upper atmosphere. In section \ref{sec:LS}, we further analyze both the cosmic muon flux and the effective atmospheric temperature using a Lomb-Scargle periodogram. In section \ref{sec:SolarAct}, we report the evidence for a long-term modulation and investigate its possible correlation with the solar cycle. In section \ref{sec:Neutron}, we report on the seasonal modulation of the cosmogenic neutron production rate in Borexino. In section \ref{sec:Concl}, we summarize our results and conclude.
\section{The Borexino Detector}
\label{sec:Borexino}
\begin{figure}[tbp]
\includegraphics[width=\textwidth]{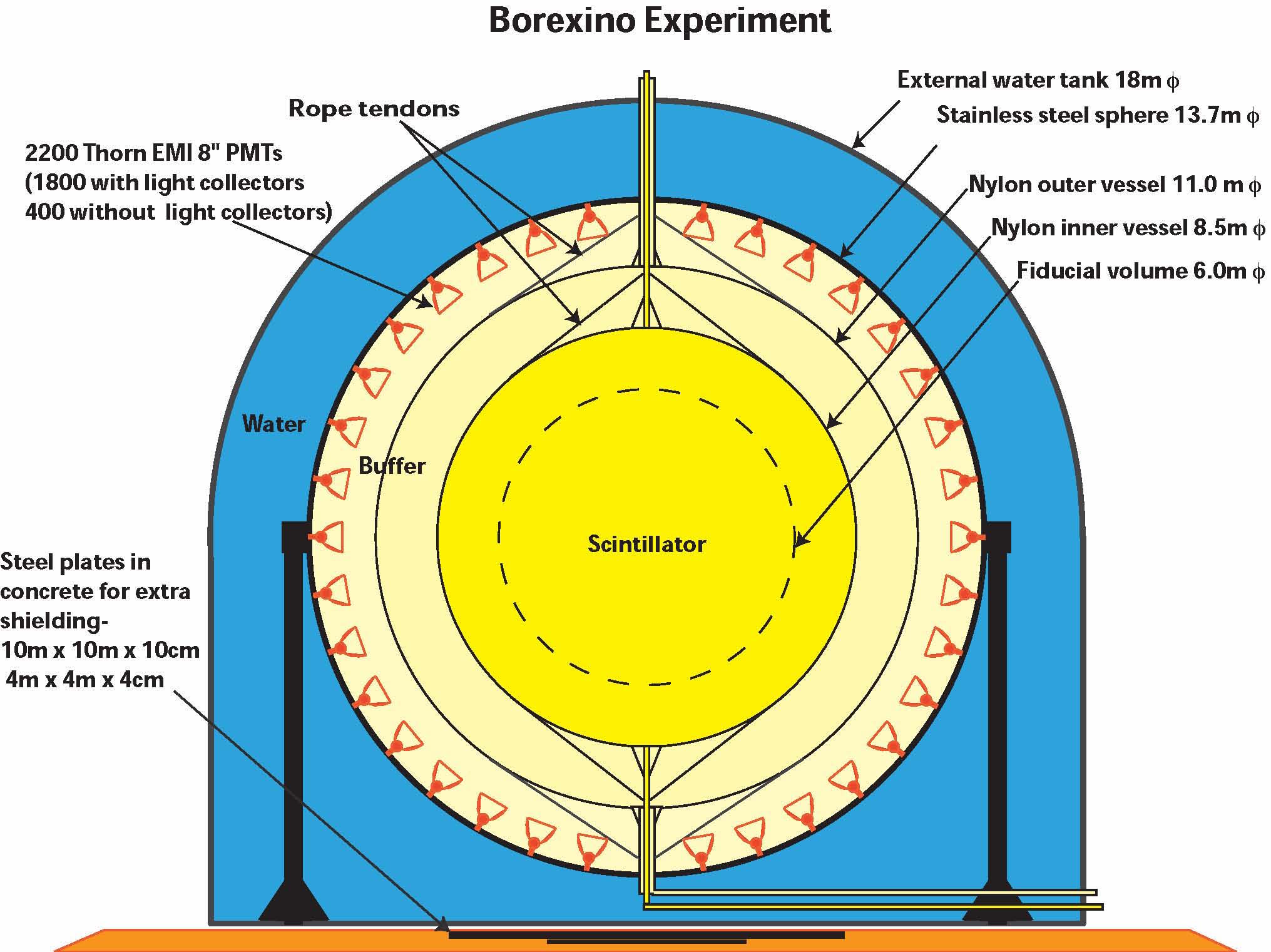}
\caption{\label{fig:BxDet}Schematic drawing of the Borexino detector.}
\end{figure}

A schematic drawing of the Borexino detector~\cite{BxDetector} is shown in figure \ref{fig:BxDet}. 
\noindent
In the present analysis, we consider muons passing through the Inner Detector (ID). It consists of a central organic scintillator target of 278\,t composed of the solvent PC (1,2,4-trimethylbenzene) doped with the wavelength shifter PPO (2,5-diphenyloxazole) at a concentration of 1.5\,g/l. The scintillator mixture is contained in a spherical and transparent nylon Inner Vessel (IV) with a diameter of 8.5\,m and a thickness of $\unit[125]{\mu m}$. To shield this central target from external $\gamma$-ray backgrounds and to absorb emanating radon, the IV is surrounded by two layers of buffer liquid in which the light quencher DMP (dimethylphthalate) is added to the scintillator solvent. A Stainless Steel Sphere (SSS) of 13.7\,m diameter holding 2212 inward-facing 8'' E.T.L. 9351 photomultiplier tubes (PMTs) that detect the scintillation light caused by particle interactions in the central region completes the ID. The ID is embedded in a steel dome of 18\,m diameter and 16.9\,m height that is filled with 2.1\,kt of ultra-pure water. Through the instrumentation of the outer surface of the SSS and the floor of the water tank with 208 PMTs, this Outer Detector (OD) provides an extremely efficient detection and tracking of cosmic muons via the Cherenkov light that is emitted during their passage through the water~\cite{BxMuon}. 
\section{Seasonal Modulation of the Cosmic Muon Flux}
\label{sec:Seas}
\begin{figure}[tbp]
\includegraphics[width = \textwidth]{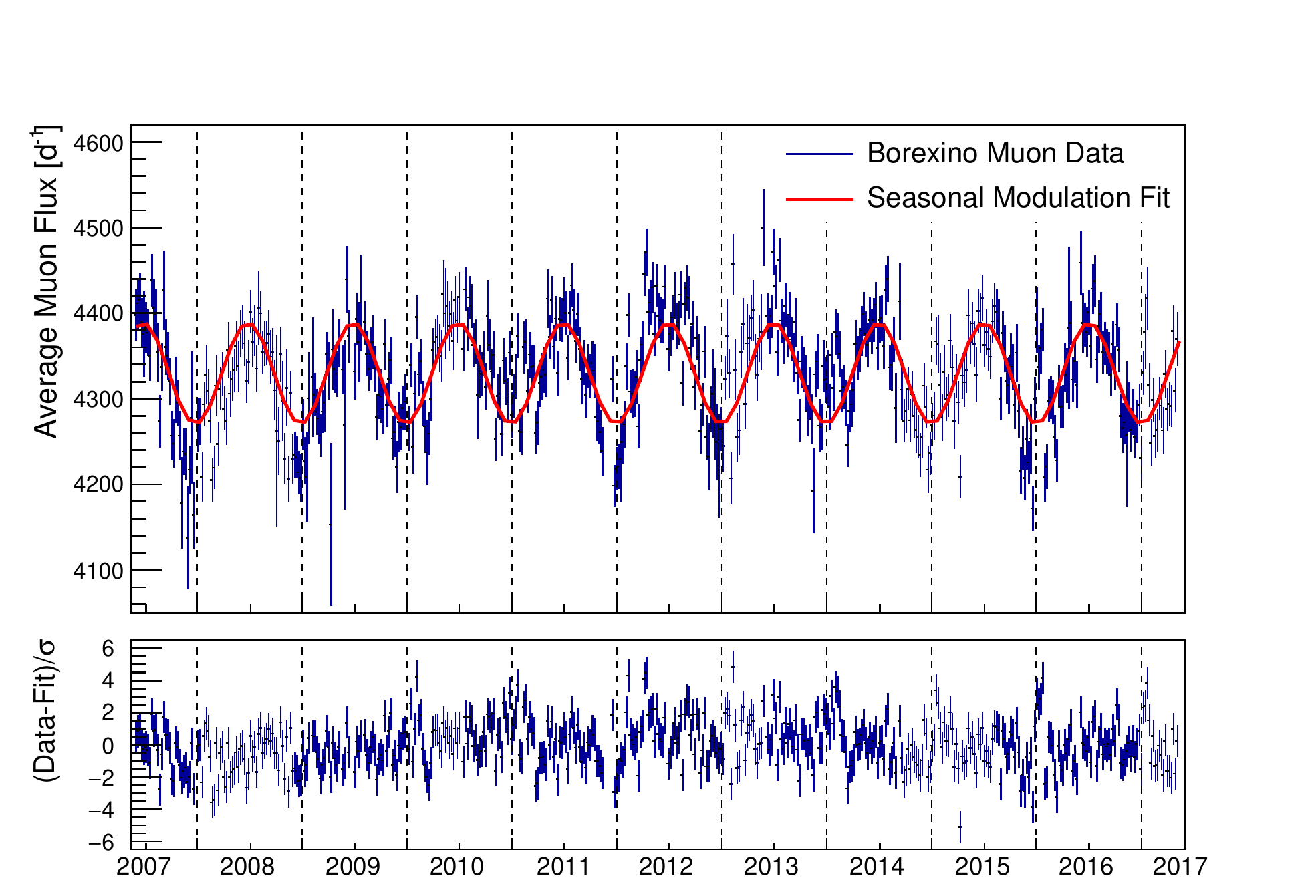}
\caption{\label{fig:MuonFlux}Cosmic muon flux measured by Borexino as a function of time. The red line depicts a sinusoidal fit to the data. The lower panel shows the residuals $\mathrm{(Data-Fit)}/\sigma$. The data are shown in weekly bins.}
\end{figure}

\begin{figure}[tbp]
\includegraphics[width = \textwidth]{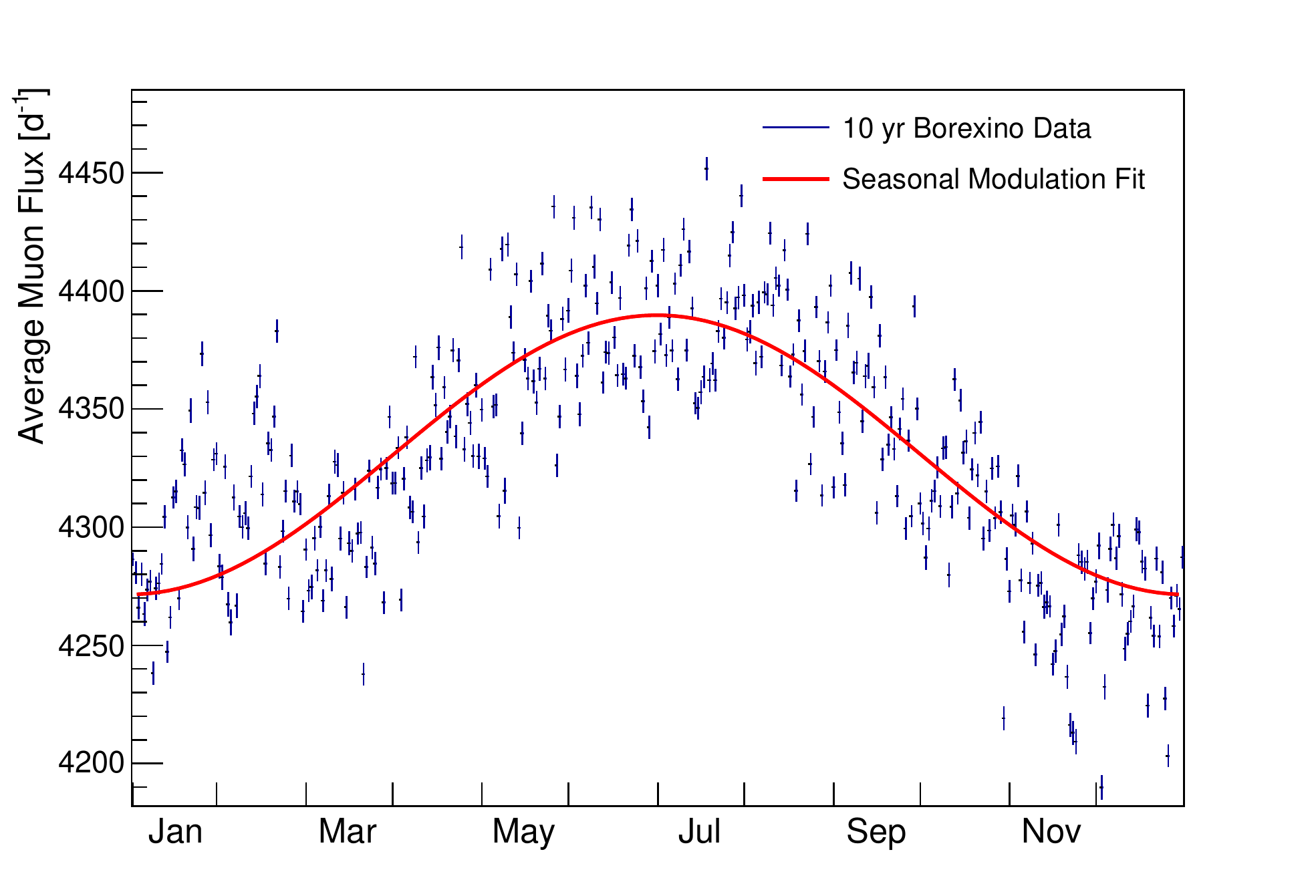}
\caption{\label{fig:Folded}Cosmic muon flux measured by Borexino in ten years folded to one year in a daily binning. The red line depicts a sinusoidal fit to the data with the period fixed to one year.}
\end{figure}

The upper atmosphere is affected by seasonal temperature variations that alter the mean free path of the muon-producing mesons at the relevant production heights. These fluctuations are expected to be mirrored in a seasonal modulation of the underground muon flux since the high energies necessary for muons to pass through the rock overburden require that the parent mesons decay in flight without any former virtual interaction.\\
The present analysis is based on ten years of Borexino data acquired between the 16$^\mathrm{th}$ of May 2007 and the 15$^\mathrm{th}$ of May 2017. Besides cosmic muons, the CERN Neutrino to Gran Sasso (CNGS) beam~\cite{CNGS} that was operational between 2008 and 2012 introduced muon events in the Borexino detector~\cite{BxSpeed}. These events have been carefully removed from the data sample via a comparison of the event time at Borexino and the beam extraction times as in~\cite{BxMuon}. To prevent statistical instabilities in the data sample, only data acquired on $3218$ days for which a minimum detector livetime of eight hours was provided are considered. Besides a phase in 2010 and 2011 during which the liquid scintillator target underwent further purification, no prolonged downtime of the detector is present in the data set.\\
Borexino features three different methods for muon identification, two of which rely on the detection of the Cherenkov light generated in the OD. The Muon Trigger Flag (MTF) is set if a trigger is issued in the OD when the detected Cherenkov light surpasses a threshold value. The Muon Clustering Flag algorithm (MCF) searches for clusters in the OD PMT hit pattern. Further, muons can be identified via their pulse shape in the ID (IDF). The mean detection efficiencies have been measured to be 0.9925(2), 0.9928(2), and 0.9890(1), respectively, and were found to remain stable. For details on the muon identification methods and the calculation of the efficiencies, we refer to~\cite{BxMuon}.\\
In the present analysis, we define muons as events that are identified by the MCF. To account for small fluctuations of the muon identification efficiency, we estimate this efficiency for each bin and correct the measured muon rate. 
We discard events that do not trigger the ID to select tracks penetrating both the ID and OD volumes. Thus, the relevant detector cross section is 146\,m$^2$ as given by the radius of the SSS, independent of the incident angle of the muon. The resulting effective exposure of the data set is $\sim 4.2\cdot 10^5\,\mathrm{m}^2\cdot \mathrm{d}$, in which  $\sim 1.2 \cdot 10^7$ muons were detected.\\
Most of the muons arriving at the Borexino detector are produced in decays of kaons and pions in the upper atmosphere. In the stratosphere, temperature modulations mainly occur on the scale of seasons, while short-term weather phenomena usually only affect the temperature of the troposphere, with the exception of stratospheric warmings that may lead to extreme temperature increases in the polar stratosphere during winter~\cite{StratWarm}. Since the higher temperature in summer lowers the average density of the atmosphere, the probability that the muon-producing mesons decay in flight before their first virtual interaction is increased due to their longer mean free paths. Only muons produced in these decays obtain enough energy to penetrate the rock coverage and reach the Borexino detector. As a consequence, the cosmic muon flux as measured by Borexino is expected to follow the modulation of the atmospheric temperature.\\
At first order, the muon flux $I_\mu(t)$ may be described by a simple sinusoidal behavior as
\begin{equation}\label{eq:Seas}
I_\mu(t) = I_\mu^0+\delta I_\mu \cos \left(\frac{2\pi}{T}(t-t_0)\right)
\end{equation}
with $ I_\mu^0$ the mean muon flux, $\delta I_\mu$ the modulation amplitude, $T$ the period, and $t_0$ the phase. Short- or long-term effects are expected to perturb the ideal seasonal modulation. Moreover, temperature and flux maxima and minima will occur at different dates in successive years.\\
The cosmic muon flux measured with Borexino is shown in figure \ref{fig:MuonFlux} together with a fit according to eq.~\ref{eq:Seas}. For better visibility, the measured average muon flux per day is shown in weekly bins while the presented results are inferred applying a fit to the muon flux in a daily binning. The lower panel shows the residuals $\mathrm{(Data-Fit)}/\sigma$.
\noindent
We measure an average muon rate $R_\mu^0 =  \unit[(4329.1 \pm 1.3)]{d^{-1}}$ in the Borexino ID after correcting for the efficiency, which corresponds to a mean muon flux $I_\mu^0= \unit[(3.432 \pm 0.001)\cdot 10^{-4}]{m^{-2}s^{-1}}$ in the LNGS. The amplitude of the clearly discernible modulation is $\delta I_\mu = \unit[(58.9 \pm 1.9)]{d^{-1}} = (1.36\pm 0.04)\%$ and we measure a period $T=\unit[(366.3 \pm 0.6)]{d}$ and a phase $t_0 = \unit[(174.8 \pm 3.8)]{d}$. This corresponds to a first flux maximum on the 25th of June 2007. The statistical uncertainties of the parameters are given and the reduced $\chi^2$ of the fit is $\chi^2 / \mathrm{NDF} = 3921 / 3214$. Here, we consider only the leading seasonal modulation of the muon flux and subleading long- or short-term effects are not accounted for in the fit function. The presence of a secondary long-term modulation that may be guessed in the residuals is investigated in sections \ref{sec:LS} and \ref{sec:SolarAct}.\\
To further analyze the phase of the seasonal modulation, we project the data to one year and fit again accordingly to eq. \ref{eq:Seas}. The period is fixed to one year as shown in figure \ref{fig:Folded}.
\noindent
While we obtain unchanged results on the mean muon flux and the amplitude of the modulation, the phase of the strictly seasonal modulation is found to be $t_0 = \unit[(181.7 \pm 0.4)]{d}$, corresponding to a maximum on the $1^\mathrm{st}$ of July. We consider this as our final estimate of the phase of the seasonal modulation. Especially in winter and spring, clear deviations from the sinusoidal assumption of the fit may be observed that can be attributed to a more turbulent environment of the upper atmosphere due to, e.g., stratospheric warmings~\cite{StratWarm}. Thus, the reduced $\chi^2$ of the fit is $\chi^2 / \mathrm{NDF} =13702 / 362$. To check the result, we selected a sample of muons as identified by the MTF and performed the same analysis steps. Consistent results were obtained and we conclude that no systematic effects based on the muon definition are introduced.\\
The flux of cosmic muons and the seasonal modulation have formerly been investigated by several experiments located at the LNGS, namely by MACRO~\cite{MACRO}, LVD~\cite{LVD,LVDII}, GERDA~\cite{GERDA}, and Borexino~\cite{BxFlux}. The results are summarized and compared to the present analysis in table \ref{tab:SeasMod}.
\begin{table}[tbp]
\centering
\begin{footnotesize}
\begin{tabular}{|c|c|c|c|c|c|c|}\hline
Experiment & Borexino & Borexino I & GERDA & MACRO & LVD I & LVD II \\
& (This Work) & \cite{BxFlux} & \cite{GERDA} & \cite{MACRO_tab} & \cite{LVD} & \cite{LVDII}\\
 \hline
Location & Hall C & Hall C & Hall A & Hall B &  Hall A & Hall A \\
Time & 2007-2017 & 2007-2011 & 2010-2013 & 1991-1997 & 2001-2008 & 1992-2016 \\ \hline
Rate  & & & & & &\\
$[\unit{10^{-4}m^{-2}s^{-1}}]$& $3.432 \pm 0.001$ & $3.41 \pm 0.01$ & $3.47 \pm 0.07$ & $3.22 \pm 0.08$ & $3.31 \pm 0.03$ & $3.3332 \pm 0.0005$ \\ \hline
Amplitude & & & & & & \\
$[\unit{10^{-6}m^{-2}s^{-1}}]$ & $4.7 \pm 0.2$ & $4.4 \pm 0.2$ & $4.72 \pm 0.33$ & -- &$5.0 \pm 0.2$ & $5.2 \pm 0.3$ \\ \hline
Amplitude & & & & & & \\
(\%) & $1.36 \pm 0.04$ & $1.29 \pm 0.07$ & $1.36 \pm 0.07$ & -- & $1.51 \pm 0.03$ & $1.56 \pm0.01 $\\ \hline
Period & & & & & & \\
$[\unit{d}]$& $366.3 \pm 0.6$ & $366 \pm 3$ & -- & -- & $367 \pm 15$ & $365.1 \pm 0.2$ \\ \hline
Phase & & & & & & \\
$[\unit{d}]$& $181.7 \pm 0.4$ & $179 \pm 3$ & $191 \pm 4$ & -- & $185 \pm 15$ & $187 \pm 3$ \\
\hline
\end{tabular}
\end{footnotesize}
\caption{\label{tab:SeasMod}Results of the cosmic muon flux modulation from Borexino compared to further measurements carried out at the LNGS. The values of the phase of the seasonal modulation were inferred via sinusoidal fits with the period fixed to one year by all experiments.}
\end{table}\noindent
The LNGS consist of three experimental halls labelled A, B, and C. Borexino reports a higher rate with respect to MACRO and LVD but a lower rate with respect to GERDA. Since the measurements were performed at the LNGS during different time epochs, the mean muon flux may be affected by variations of the mean temperature or by a long-term modulation of the cosmic muon flux. Further, unlike Borexino, the acceptance of the other experiments carried out at the LNGS contains a dependence on the incident angle of the muons that must be carefully modelled. The seasonal modulation is found by all experiments and the phases agree well with that determined in the present work. Only GERDA reports a later maximum of the cosmic muon flux but their analysis is based on three years of data only.
\section{Atmospheric Model and Effective Atmospheric Temperature}
\label{sec:AtmoMod}
Since the mesons, and consequently also the muons from their decays, are produced at various heights in the atmosphere, it is extremely difficult to determine the point in its temperature distribution where an individual muon was produced. In order to investigate the correlation between fluctuations of the atmospheric temperature and the cosmic muon flux observed underground, the atmosphere is modelled as an isothermal meson-producing entity with an effective temperature $T_\mathrm{eff}$~\cite{Barrett}. $T_\mathrm{eff}$ is defined as the temperature of an isothermal atmosphere that produces the same meson intensities as the actual atmosphere. Properly chosen weighting factors must be assigned to the corresponding depth levels accounting for the physics that determine the meson and muon production.\\
A common parametrization is given by~\cite{MINOS,MINOS_RKPI}
\begin{equation}\label{eq:teff}
\begin{split}
T_\mathrm{eff} &= \frac{\int_0^\infty \mathrm{d}X\, T(X)\alpha^\pi(X)  + \int_0^\infty \mathrm{d}X\, T(X)\alpha^\mathrm{K}(X) }{\int_0^\infty \mathrm{d}X \, \alpha^\pi(X)   +   \int_0^\infty \mathrm{d}X \, \alpha^\mathrm{K}(X) }\\
&\simeq \frac{\sum_\mathrm{n=0}^N  \Delta X_\mathrm{n}\, T(X_\mathrm{n}) (W_\mathrm{n}^\pi+W_\mathrm{n}^\mathrm{K})}{\sum_\mathrm{n=0}^N  \Delta X_\mathrm{n}\ (W_\mathrm{n}^\pi+W_\mathrm{n}^\mathrm{K})},
\end{split}
\end{equation}
where the approximation considers that the temperature is measured at discrete levels $X_\mathrm{n}$. The temperature coefficients $\alpha^\pi(X)$ and $\alpha^\mathrm{K}(X)$ relate the atmospheric temperature to the muon flux considering pion and kaon contributions, respectively. These coefficients are translated into the weights $W^\pi_\mathrm{n}$ and $W^\mathrm{K}_\mathrm{n}$ via numerical integration over the atmospheric levels $\Delta X_\mathrm{n}$ to allow the approximation. The weights are defined in appendix \ref{ap:A}.\\
Figure \ref{fig:weights} shows the ten year average temperature at different pressure levels using data for the closest point to the LNGS as provided by the European Center for Medium-range Weather Forecasts (ECMWF)~\cite{ecmwf} and the assigned weights to the respective altitude levels.
\begin{figure}[tbp]
\includegraphics[width = \textwidth]{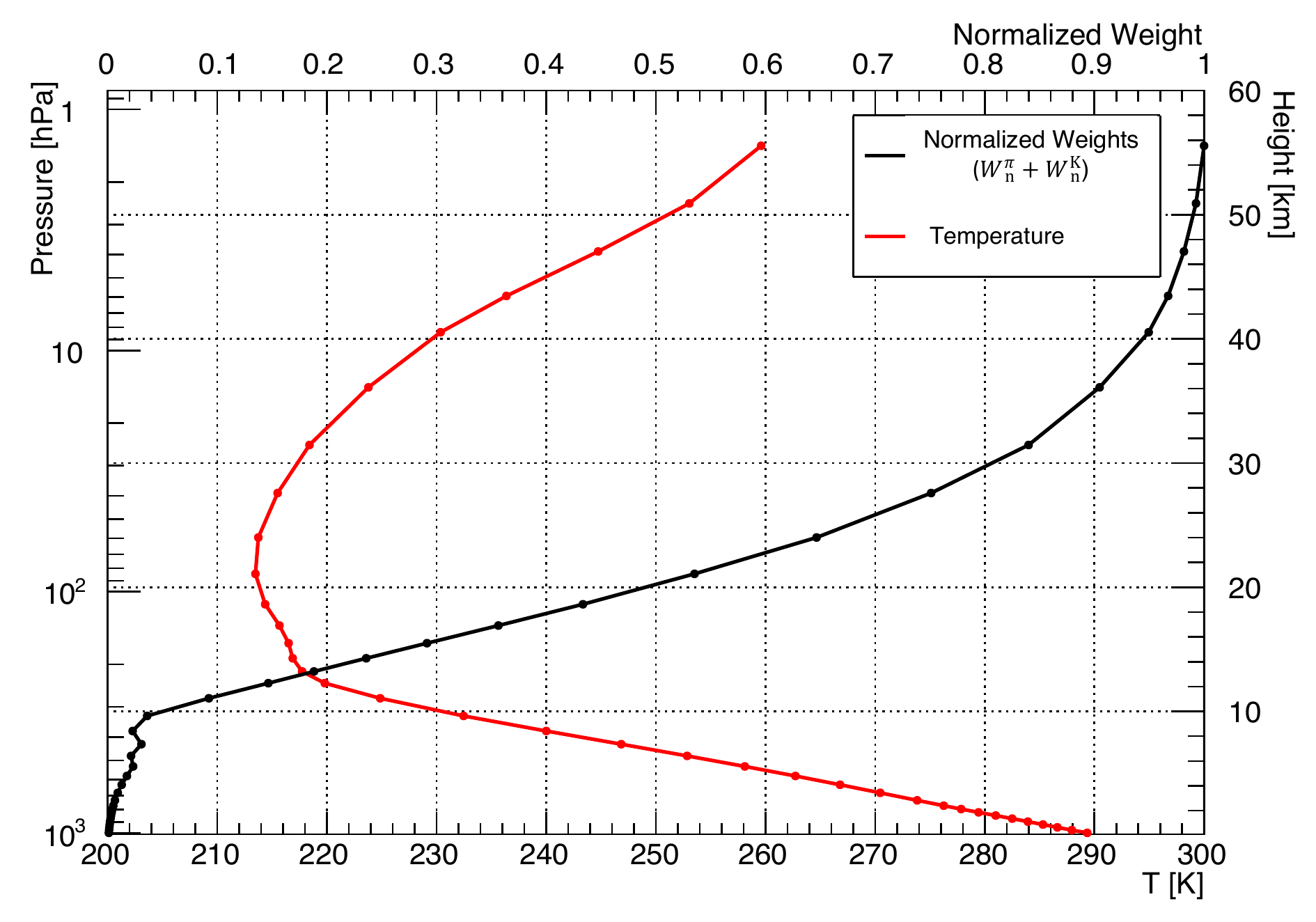}
\caption{\label{fig:weights}The ten year average temperature~\cite{ecmwf} at the location of the LNGS is shown by the red line and the normalized weighting factors $W_\mathrm{n}^\pi+W_\mathrm{n}^\mathrm{K}$ by the black line, as functions of the pressure levels. The right vertical axis shows the altitude corresponding to the pressure level on the left vertical axis.}
\end{figure}\noindent
Higher layers of the atmosphere are assigned higher weights since muons possessing sufficient energy to penetrate the rock coverage of the LNGS are mainly produced at these altitudes. On the contrary, muons produced at lower altitudes are usually less energetic and the majority will not have the threshold energy $E_\mathrm{thr}$ to reach the detector.\\
A so-called effective temperature coefficient may be defined as 
\begin{equation}\label{eq:alpha}
\alpha_\mathrm{T} = \frac{T_\mathrm{eff}}{I_\mu^0} \int_0^\infty \mathrm{d} X \, W(X),
\end{equation}
where $W(X) = W^\pi(X) +W^\mathrm{K}(X)$. Thus, fluctuations of the cosmic muon flux may be related to fluctuations of the effective temperature via
\begin{equation}\label{eq:linear_alpha}
\frac{\Delta I_\mu}{I_\mu^0} =  \alpha_\mathrm{T} \frac{\Delta T_\mathrm{eff}}{T_\mathrm{eff}}
\end{equation} 
and $\alpha_\mathrm{T}$ quantifies the correlation between these two observables as discussed in section \ref{sec:Corr}.
\section{Seasonal Modulation of the Effective Atmospheric Temperature}
\label{sec:TempMod}
\begin{figure}[tbp]
\includegraphics[width=\textwidth]{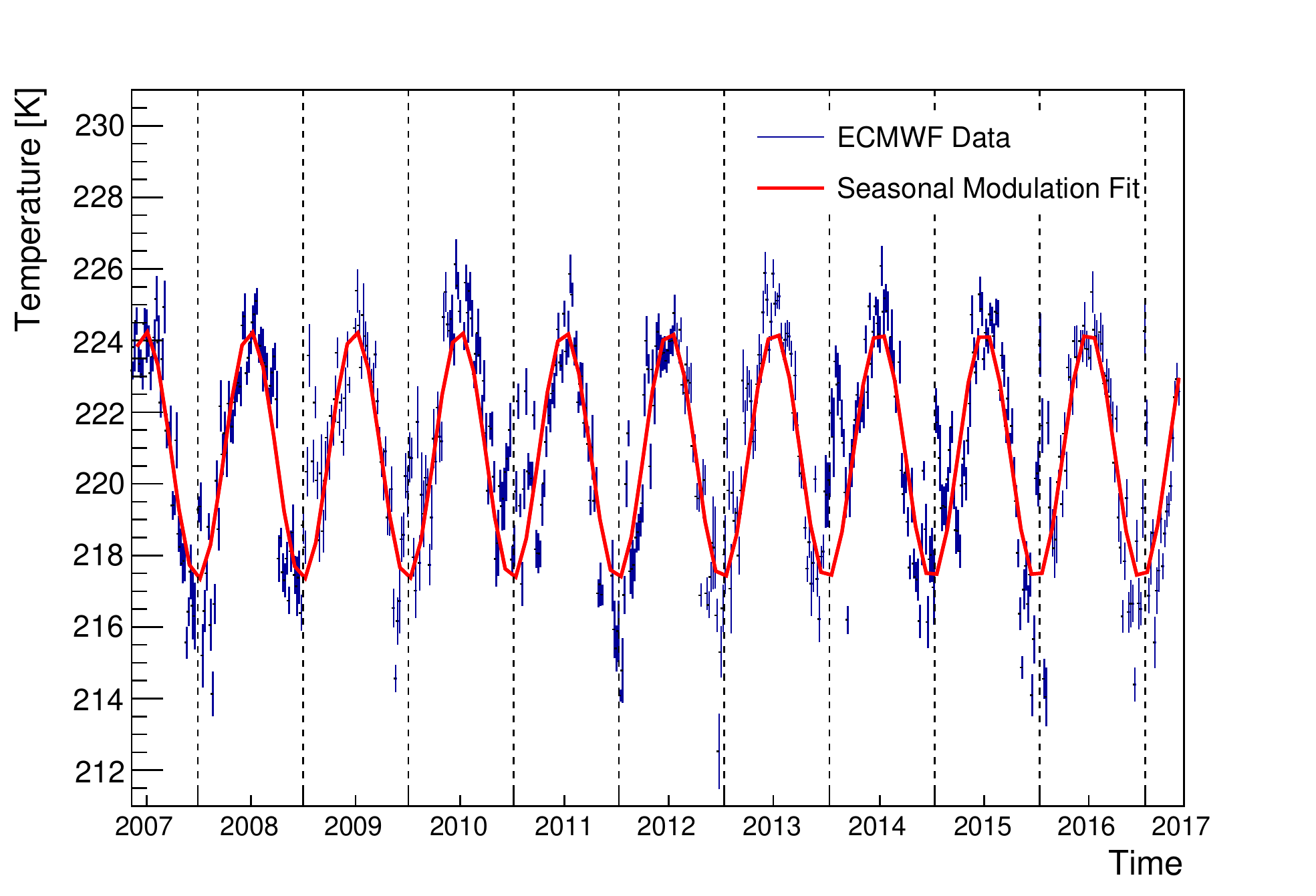}
\caption{\label{fig:TempMod}Effective atmospheric temperature computed accordingly to eq. \ref{eq:teff}. The curve shows a sinusoidal fit of the data.}
\end{figure}

To verify the correlation between the observed modulation of the cosmic muon flux and fluctuations of the atmospheric temperature, we analyze atmospheric temperature data provided by the ECMWF~\cite{ecmwf} for the time period corresponding to the muon flux measurement. This data is generated by interpolating several atmospheric observables based on different types of observations (surface measurements, satellite data, or upper air sounding) and a global atmospheric model. For this analysis, we use the temperature for the location at $42.75^{\circ}$N and $13.5^{\circ}$E, which is the closest grid point to the LNGS available. The model provides atmospheric temperature data at 37 discrete pressure levels in the range from [0-1000]\,hPa four times per day at 00.00\,h, 06.00\,h, 12.00\,h, and 18.00\,h. Based on these data, we calculated $T_\mathrm{eff}$ for each of the temperature sets based on eq. \ref{eq:teff}. The effective atmospheric temperature $T_\mathrm{eff}$ of the respective day was computed as the average of the four values calculated during the day, their variance was used to estimate the uncertainty.\\
Figure \ref{fig:TempMod} shows the mean effective atmospheric temperature in a weekly binning. Analogously to the cosmic muon flux, the modulation parameters were inferred by a fit to the data in a daily binning.
\noindent
A fit similar to eq. \ref{eq:Seas} returns an average effective atmospheric temperature $T_\mathrm{eff}^0 = \unit[(220.893 \pm 0.005)]{K}$, a modulation amplitude $\delta T_\mathrm{eff} = \unit[(3.43\pm 0.01)]{K} = (1.56 \pm 0.01)\%$, a period $\tau=\unit[(365.69 \pm 0.04)]{d}$, and a phase $t_0=\unit[(180.8 \pm 0.2)]{d}$. While period and phase of the temperature modulation clearly show the leading seasonal behavior of the effective atmospheric temperature and agree well with the results of the muon flux discussed in section \ref{sec:Seas}, the slightly higher modulation amplitude indicates that not all mesons relevant for the production of muons penetrating the LNGS rock coverage are affected by the density variations of the atmosphere. With a $\chi^2 / \mathrm{NDF} = 118460 / 3649$, a sinusoidal is only a very poor reproduction of the fine-grained temperature data. Similar to the flux of cosmic muons (see section \ref{sec:Seas}), short-term variations of the effective atmospheric temperature and, especially, additional secondary maxima in winter and spring are observed. These maxima may be ascribed to stratospheric warmings~\cite{StratWarm}. Sudden Stratospheric Warmings (SSW) sometimes even feature amplitudes comparable to the leading seasonal modulation~\cite{MINOSSSW}, as visible e.g. in winter 2016/2017.
\section{Correlation Between Muon Flux and Temperature}
\label{sec:Corr}
\begin{figure}[tbp]
\includegraphics[width = \textwidth]{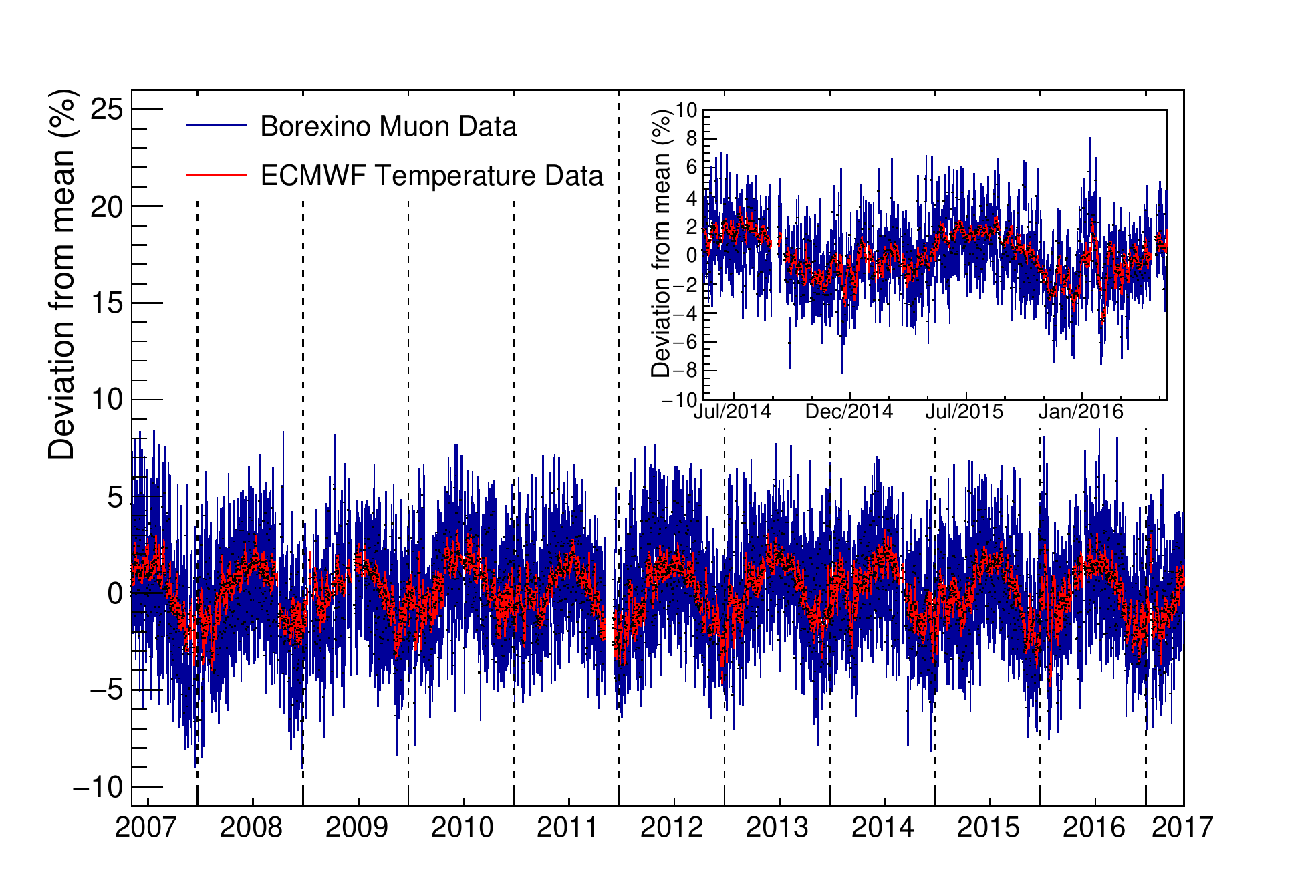}
\caption{\label{fig:TempMuon}Daily percent deviations of the cosmic muon flux and the effective atmospheric temperature from the mean in ten years of data. The insert shows a zoom for two years from May 2014 to May 2016.}
\end{figure}

\begin{figure}[tbp]
\includegraphics[width = \textwidth]{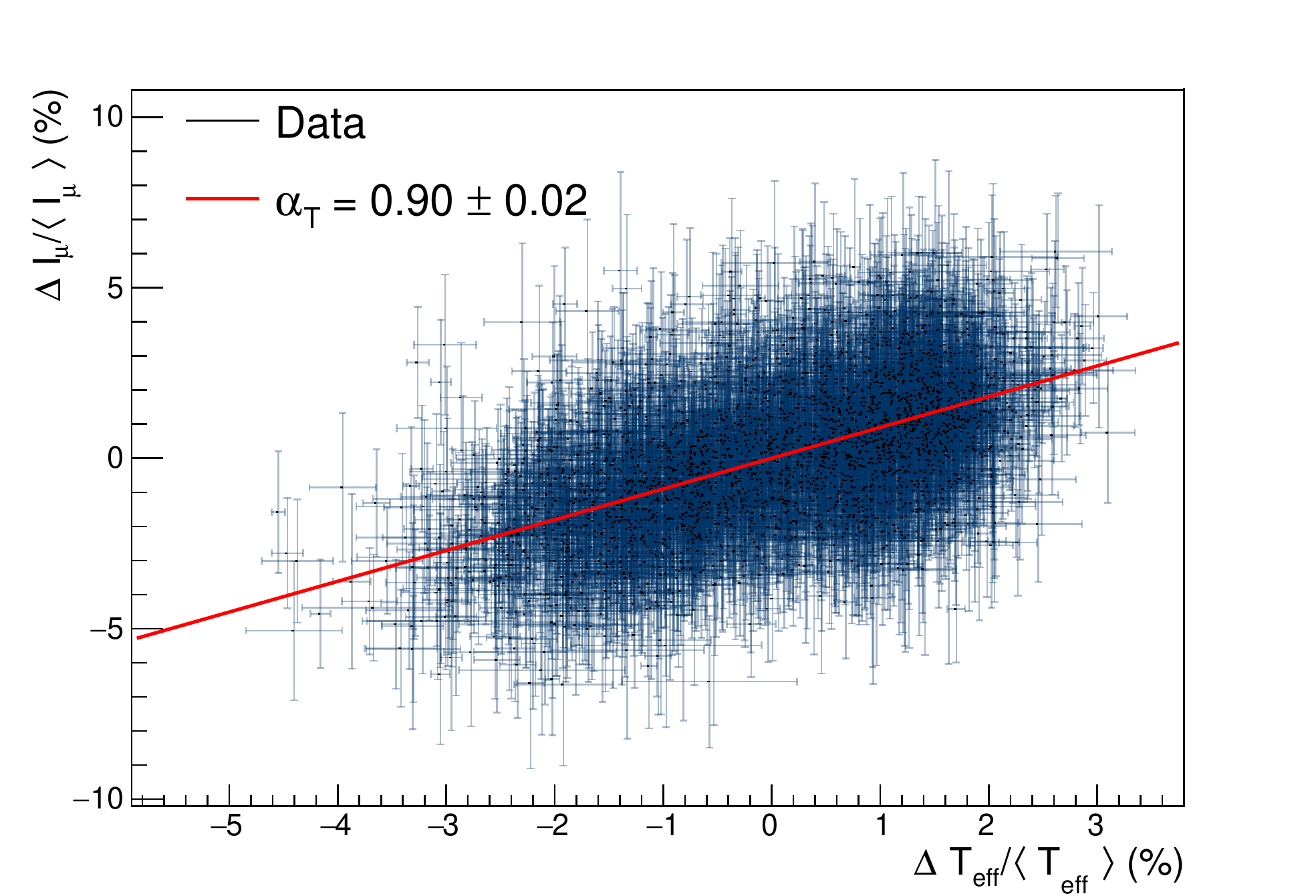}
\caption{\label{fig:corr}$\Delta I_\mu / I_\mu^0$ versus $\Delta T_\mathrm{eff}/T_\mathrm{eff}^0$ with each point corresponding to one day.}
\end{figure}

As expected, the modulation parameters inferred for the cosmic muon flux in section \ref{sec:Seas} and the effective atmospheric temperature in section \ref{sec:TempMod} point towards a correlation of the two observables. Figure \ref{fig:TempMuon} shows the measured muon flux in Borexino and the effective atmospheric temperature scaled to percent deviations from their means $I_\mu^0$ and $T_\mathrm{eff}^0$ for ten years in a daily binning. $I_\mu^0$ and $T_\mathrm{eff}^0$ were determined via sinusoidal fits to the respective data sets.
\noindent
Besides the consistency of the leading seasonal modulations of both observables, we find short-term variations of the temperature to be promptly mirrored in the underground muon flux. Exemplarily, the short-term and non-seasonal temperature increase around January 2016 generates a secondary maximum of the muon flux.\\
To quantify the correlation of the two observables, we plot $\Delta I_\mu / I_\mu^0$ versus $\Delta T_\mathrm{eff}/T_\mathrm{eff}^0$ for each day as shown in figure \ref{fig:corr}.
\noindent
Indeed, we find a positive correlation coefficient (R-value) of 0.55.\\
Based on eq. \ref{eq:linear_alpha}, we determine the effective temperature coefficient by performing a linear regression using a numerical minimization method and accounting for error bars on both axes. We obtain $\alpha_\mathrm{T}=0.90 \pm 0.02_\mathrm{stat.}$ in agreement with the former Borexino result of $\alpha_\mathrm{T}=0.93 \pm 0.04_\mathrm{stat.}$~\cite{BxFlux} but with the statistical uncertainties reduced by a factor $\sim 2$.\\ 
In order to analyze systematic uncertainties, we performed the following checks: (1) We repeated the analysis selecting muons with our alternative muon identification method MTF. An effective temperature coefficient $\alpha_\mathrm{T}(\mathrm{MTF})=0.92 \pm 0.02_\mathrm{stat.}$ is measured in agreement with the above result.
(2) We allowed for an offset in eq.~\ref{eq:linear_alpha} and fit the data. The fit provides an intercept $\alpha_0=-0.02 \pm 0.03$ consistent with zero, meaning that no obvious offsets or non-linearities are observed.
(3) We performed the analysis for a two-year moving subset. We find the result to be stable and consistent with the full data set without any fluctuations above the statistical expectations.
We conclude that any systematic uncertainty must be small compared to the statistical uncertainty obtained from the fit.\\
In table \ref{tab:alpha}, the result of this analysis is compared to several further measurements performed at the LNGS. The results agree well within their uncertainties. The GERDA experiment~\cite{GERDA} reported two values of $\alpha_\mathrm{T}$ using two different sets of temperature data. 
\begin{table}[tbp]
\centering
\begin{tabular}{|c|c|c|}\hline
Experiment & Time period & $\alpha_\mathrm{T}$ \\ \hline
Borexino (This work) & 2007-2017 & $0.90 \pm 0.02$ \\ \hline
Borexino Phase I~\cite{BxFlux} & 2007-2011 & $0.93 \pm 0.04$ \\ \hline
GERDA~\cite{GERDA} & 2010-2013 & $0.96 \pm 0.05$ \\
& & $0.91 \pm 0.05$ \\ \hline
MACRO~\cite{MACRO_tab} & 1991-1997 & $0.91 \pm 0.07$\\ \hline
LVD~\cite{LVDII}  & 1992-2016 & $0.93 \pm 0.02$ \\
\hline
\end{tabular}
\caption{\label{tab:alpha}Comparison of measurements of the effective temperature coefficient at the LNGS.}
\end{table}\noindent
The theoretical expectation of $\alpha_\mathrm{T}$ at the location of the LNGS considering muon production from both kaons and pions was formerly calculated in~\cite{BxFlux} to be $0.92\pm 0.02$ assuming $\langle E_\mathrm{thr} \cos \theta \rangle =\unit[1.833]{TeV}$ based on~\cite{MINOS_RKPI}. With the threshold energy $\langle E_\mathrm{thr} \cos \theta \rangle=\unit[(1.34 \pm 0.18)]{TeV}$ estimated in this paper (see section \ref{sec:RKPI}), the expectation is $\alpha_\mathrm{T} = 0.893\pm 0.015$. Hence, our measurement is still in agreement with both estimations.
\section{Atmospheric Kaon-to-Pion Production Ratio}
\label{sec:RKPI}
\begin{figure}
\includegraphics[width = \textwidth]{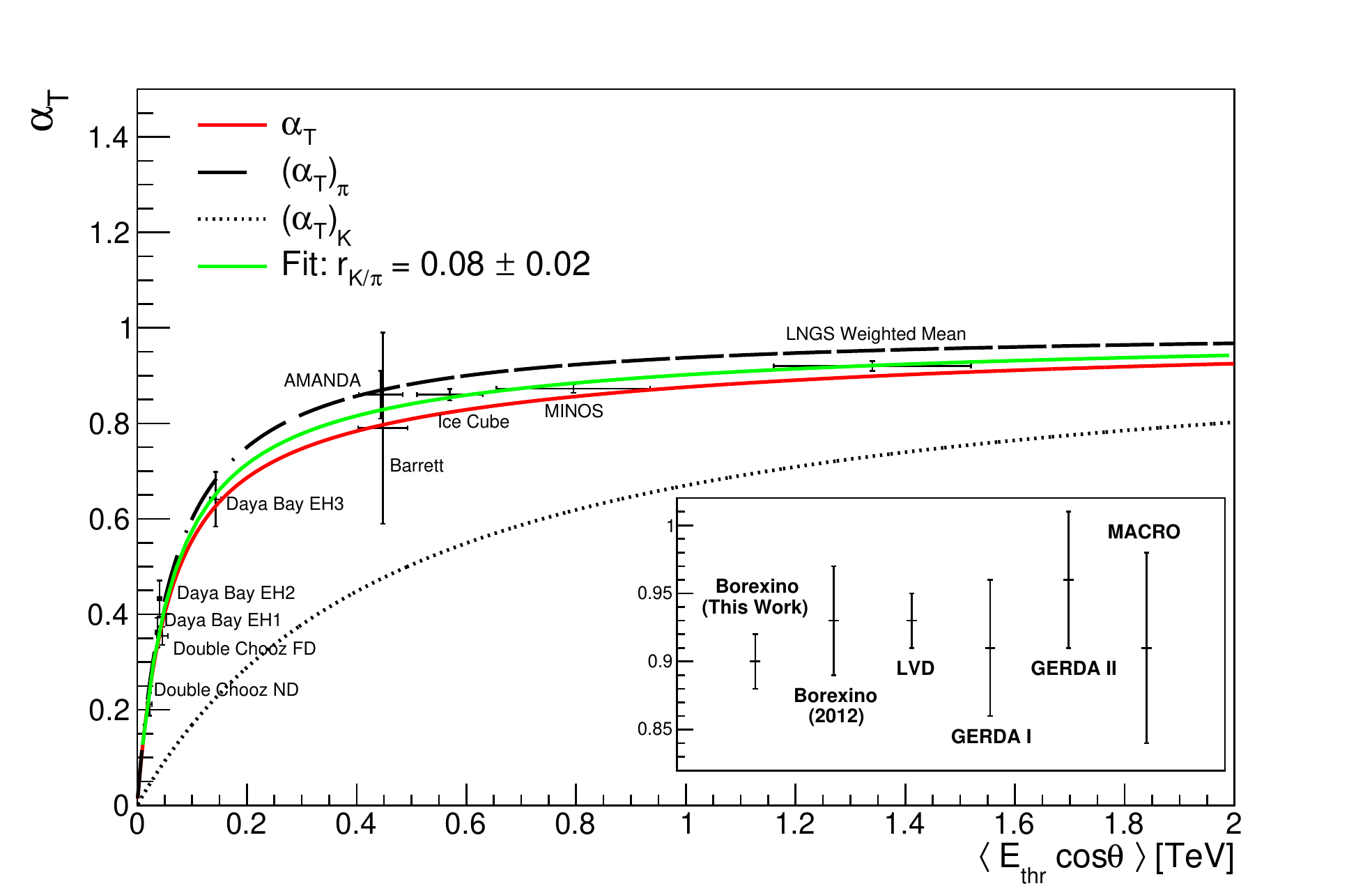}
\caption{\label{fig:alpha_comp}Measurements of the effective temperature coefficient $\alpha_\mathrm{T}$ at varying $\langle E_\mathrm{thr} \cos \theta \rangle$. The curves indicate the expected $\alpha_\mathrm{T}$ for different assumptions of $ r_{\mathrm{K}/\pi}$, with the green curve showing a fit of the measurements according to eq. \ref{eq:alpha_MC}. The insert shows the result of the present work compared to measurements from other LNGS-based experiments.}
\end{figure}

\begin{figure}[tbp]
\includegraphics[width = \textwidth]{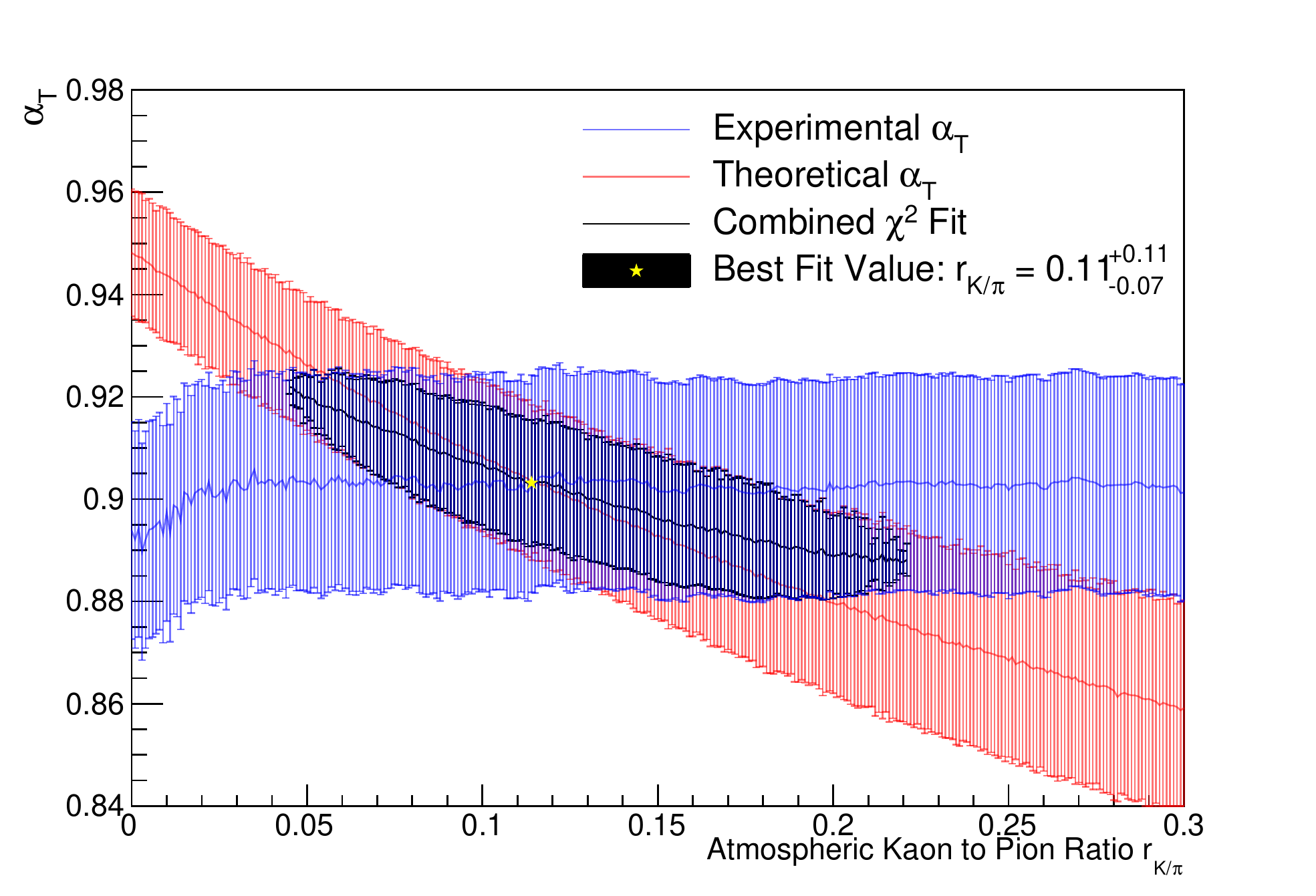}
\caption{\label{fig:RKPI}Measured value of $\alpha_\mathrm{T}$ in blue and theoretical prediction in red as functions of $r_{\mathrm{K}/\pi}$. The black region indicates the $1\sigma$ contour of the intersection region of $r_{\mathrm{K/\pi}}=0.11^{+0.11}_{-0.07}$ around the best fit value marked by the yellow star.}
\end{figure}

\begin{figure}
\includegraphics[width = \textwidth]{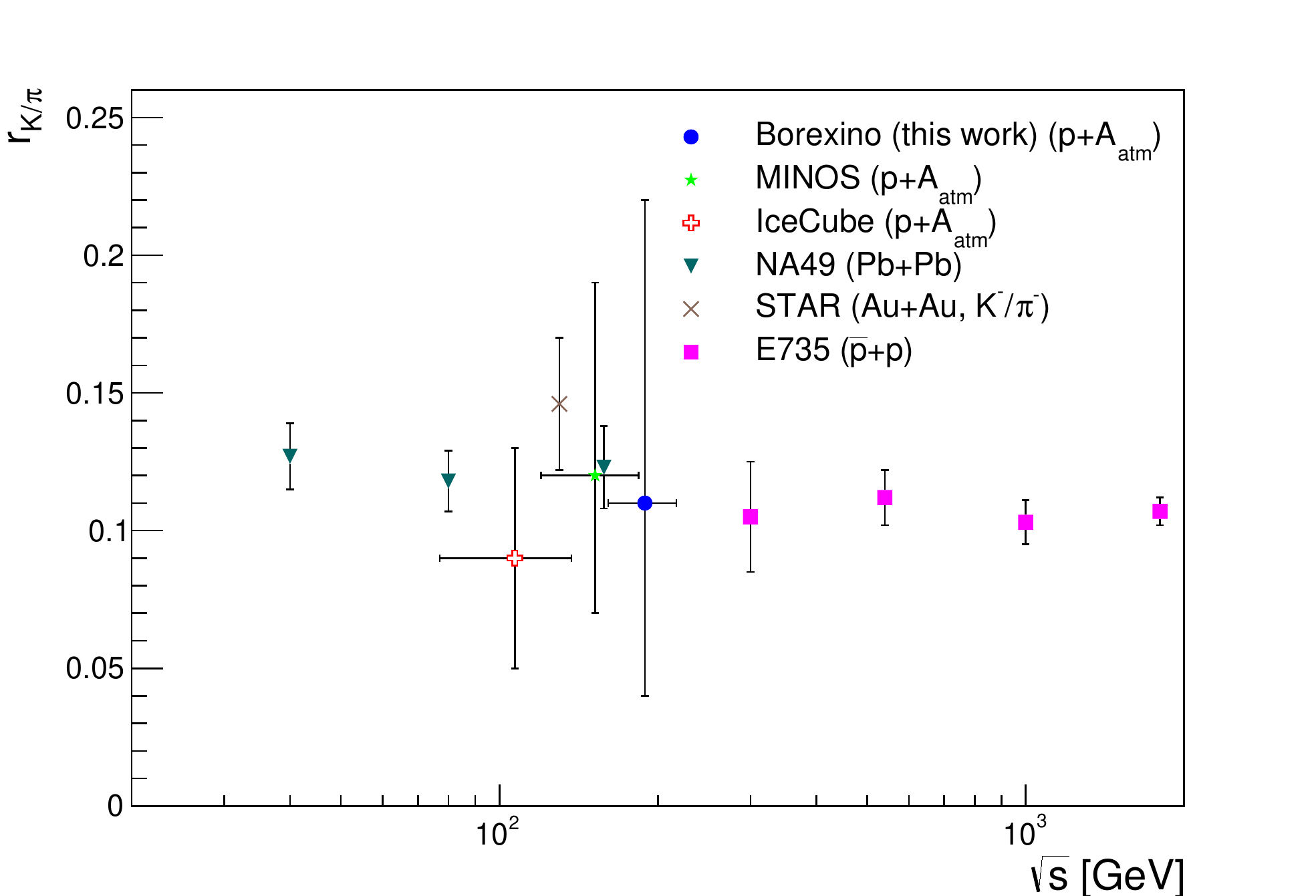}
\caption{\label{fig:comp_rkpi}Comparison of several measurements of the kaon-to-pion production ratio. The STAR measurement was performed using Au+Au collisions at RHIC~\cite{AdlerAuAuRKPI}, the NA49 using Pb+Pb collisions at SPS~\cite{AfanasievPbPbRKPI}, and the E735 using $\mathrm{\bar{p}+p}$ collisions at Tevatron~\cite{ppbarcoll}. The MINOS~\cite{MINOS}, IceCube~\cite{IceCube}, and Borexino measurements were performed indirectly via a measurement of the effective temperature coefficient.}
\end{figure}

Since kaons and pions are affected differently by atmospheric temperature variations due to their distinct properties like mass, lifetime, or attenuation length, the strength of the correlation between the cosmic muon flux underground and the atmospheric temperature depends on the production ratio of kaons and pions in the atmosphere. In the following, we infer an indirect measurement of the atmospheric kaon-to-pion production ratio based on the measurement of the effective temperature coefficient reported in section \ref{sec:Corr}.\\
For a properly weighted temperature distribution, the effective temperature coefficient $\alpha_\mathrm{T}$ is theoretically predicted to be~\cite{Barrett}
\begin{equation}\label{eq:alphatheo}
\alpha_\mathrm{T} = \frac{T}{I_\mu^0} \frac{\partial I_\mu}{\partial T}
\end{equation}
with $T$ being the temperature. The differential muon spectrum at the surface may be parametrized as~\cite{Gaisser}
\begin{equation}\label{eq:muonspec}
\frac{\mathrm{d}I_\mu}{\mathrm{d}E_\mu} \simeq A \times E_\mu^{-(\gamma+1)} \left(  \frac{1}{1+1.1E_\mu \cos \theta / \epsilon_\pi} +   \frac{0.38\cdot r_{\mathrm{K}/\pi}}{1+1.1E_\mu \cos \theta / \epsilon_\mathrm{K}} \right)
\end{equation}
with $r_{\mathrm{K}/\pi}$ the atmospheric kaon-to-pion ratio, $\theta$ the zenith angle, $\gamma=1.78 \pm 0.05$~\cite{LVDVert} the muon spectral index, and $\epsilon_\pi = \unit[(114 \pm 3)]{GeV}$ and $\epsilon_\mathrm{K} = \unit[(851\pm 14)]{GeV}$~\cite{MINOS} the critical pion and kaon energies, respectively. The critical meson energy separates the decay from the interaction regime: mesons with an energy below this energy are more likely to decay, while mesons with a higher energy most probably interact in the atmosphere before decaying.\\
As shown in~\cite{Barrett}, eq. \ref{eq:alphatheo} may be transformed into
\begin{equation}\label{eq:alpha_trans}
\alpha_\mathrm{T}=-\frac{E_\mathrm{thr}}{I_\mu^0} \frac{\partial I_\mu}{\partial E_\mathrm{thr}} - \gamma
\end{equation}
with the threshold energy $E_\mathrm{thr}$. The muon intensity underground may be approximated for the muon surface spectrum described by eq. \ref{eq:muonspec} as~\cite{Barrett,MACRO}
\begin{equation}
I_\mu \simeq B \times E_\mathrm{thr}^{-\gamma} \left[\frac{1}{\gamma +(\gamma +1)\,1.1 \langle E_\mathrm{thr}\cos\theta\rangle / \epsilon_\pi} +  \frac{0.38 \cdot r_{\mathrm{K}/\pi}}{\gamma +(\gamma +1)\,1.1 \langle E_\mathrm{thr}\cos\theta \rangle/ \epsilon_\mathrm{K}}   \right].
\end{equation}
With this approximation, the predicted $\alpha_\mathrm{T}$ may be calculated as
\begin{equation}\label{eq:alpha_MC}
\alpha_\mathrm{T} = \frac{1}{D_\pi} \frac{1/\epsilon_\mathrm{K}+A_\mathrm{K}(D_\pi/D_\mathrm{K})^2/\epsilon_\pi}{1/\epsilon_\mathrm{K}+A_\mathrm{K}(D_\pi/D_\mathrm{K})/\epsilon_\pi},
\end{equation}
with
\begin{equation}\label{eq:DPI}
D_{\pi,\mathrm{K}} \equiv \frac{\gamma}{\gamma+1}\, \frac{\epsilon_{\pi,\mathrm{K}}}{1.1\langle  E_\mathrm{thr}\cos \theta \rangle} +1
\end{equation}
and $A_\mathrm{K}=0.38 \times r_{\mathrm{K}/\pi}$ describing the kaon contribution to the cosmic muon flux~\cite{MINOS}. $E_\mathrm{thr}\cos(\theta)$ is the product of the threshold energy for a muon arriving from a zenith angle $\theta$ at the detector and the cosine of this angle. The mean value of this product allows to properly parametrize and compare the depths of various underground sites taking into account that the threshold energy is direction-dependent due to the shape of the respective rock overburden.\\
Figure \ref{fig:alpha_comp} shows the weighted mean of $\alpha_\mathrm{T}$ for measurements performed at the LNGS together with measurements at other underground laboratories from Barrett~\cite{Barrett}, IceCube~\cite{IceCube}, \mbox{MINOS}~\cite{MINOS}, Double Chooz~\cite{DoubleChooz}, Daya Bay~\cite{DAYA}, and AMANDA~\cite{AMANDA}. The experimental results are plotted as a function of $\langle E_\mathrm{thr} \cos \theta \rangle$, which is the parameter on which $\alpha_\mathrm{T}$ explicitly depends (eq. \ref{eq:alpha_MC}-\ref{eq:DPI}). The insert shows the LNGS based measurements from MACRO~\cite{MACRO}, LVD~\cite{LVDII}, GERDA~\cite{GERDA}, and the two Borexino measurements from 2012~\cite{BxFlux} and from this work.
\noindent
For the LNGS, a value of $\langle E_\mathrm{thr} \cos \theta \rangle = \unit[(1.34\pm0.18)]{TeV}$ has been calculated based on a Monte Carlo simulation (see below). The red line shows the expected $\alpha_\mathrm{T}$ as a function of $\langle E_\mathrm{thr} \cos \theta \rangle$ considering muon production using the literature value of the atmospheric kaon-to-pion ratio of $ r_{\mathrm{K}/\pi} = 0.149 \pm 0.06$~\cite{RKPILit}, the dashed and dotted lines illustrate the extreme cases of pure pion or pure kaon production, respectively. The green line indicates the result of a fit to the measurements according to eq. \ref{eq:alpha_MC} with $r_{\mathrm{K}/\pi}$ as a free parameter. We obtain $r_{\mathrm{K}/\pi}=0.08 \pm 0.02_\mathrm{stat.}$ at a $\chi^2/\mathrm{NDF}=5/9$. However, note that systematic uncertainties like the exact value of $\langle E_\mathrm{thr} \cos \theta \rangle$ for the respective experimental sites are not fully determined and this result is only indicative. Also, the measured values of $\alpha_\mathrm{T}$ depend on the assumed kaon-to-pion ratio since this quantity is included in the computation of $T_\mathrm{eff}$. We do not take into account this inter-value dependency here.\\
The value of $r_{\mathrm{K}/\pi}$ can also be inferred indirectly from the combination of a theoretical calculation of $\alpha_\mathrm{T}$ with the measurement from Borexino. In this case, no further experimental data has to be included. We performed a Monte Carlo simulation to calculate the expected value of $\langle \alpha_\mathrm{T} \rangle$ at the location of the LNGS depending on $r_{\mathrm{K}/\pi}$. For muons with an energy $E_{\mathrm{\mu}} \gg \epsilon_{\mathrm{\pi}}$, which is true for the muons arriving at the LNGS, the zenith angle distribution is best described by $\sec \theta$ for $\theta < 70^\circ$ instead of the usual $\cos^2 \theta$~\cite{pdg}.  We generated a toy Monte Carlo set of muons by randomly drawing a zenith angle from this distribution and an energy from the distribution given by eq. \ref{eq:muonspec} for the respective $\theta$. Moreover, a random azimuth angle $\phi$ was selected and the rock coverage $D(\theta, \phi)$ for muons arriving from this direction was calculated based on an altitude profile of the Gran Sasso mountains obtained from the Google Maps Elevation API~\cite{google} and the density of the Gran Sasso rock of $\rho = \unit[(2.71 \pm 0.05)]{g/cm^3}$~\cite{gsdens}. We converted this into a direction dependent threshold energy $E_\mathrm{thr}(\theta, \phi)$ for surface muons to reach the LNGS using the energy loss formula given in~\cite{pdg} with fixed parameters. For $r_{\mathrm{K}/\pi}$ values increasing from 0 to 0.3 in steps of 0.01, we calculate the corresponding mean value of the effective temperature coefficient $\langle \alpha_\mathrm{T} \rangle$ for samples of 10000 muons with $E_\mu>E_\mathrm{thr}(\theta, \phi)$.\\
To check our results, we performed the same calculations using the depth and zenith angle distributions of muons arriving at the LNGS predicted by the MUSUN (MUon Simulations UNderground)~\cite{MUSUN} simulation code for this location. We obtain a close agreement between the two simulations with a mean difference $\langle \Delta \alpha_\mathrm{T}\rangle=3.6\cdot 10^{-4}$. Additionally, we compared the zenith angle distribution predicted by our simulation with the measured distribution and found them to be in good agreement.
To estimate the systematic uncertainty of $\langle \alpha_\mathrm{T} \rangle$, we varied the input parameters of the simulation. We considered contributions from a $5\%$ uncertainty of the altitude profile, the uncertainty of the measured rock density of the Gran Sasso rock of $\unit[0.05]{g/cm^3}$, the uncertainty of the measurement of the muon spectral index of $0.05$~\cite{LVDVert}, the uncertainties of the critical meson energies $\Delta \epsilon_\pi=\unit[3]{GeV}$ and $\Delta \epsilon_\mathrm{K}= \unit[14]{GeV}$, and a $10\%$ uncertainty of the drawn zenith angle. For the combined systematic uncertainty of $\langle \alpha_\mathrm{T} \rangle$, we found $\sim 0.015$. However, the strength of several of the contributions coming from the above factors depends on $r_{\mathrm{K}/\pi}$. In particular, the larger uncertainty of $\epsilon_\mathrm{K}$ compared to $\epsilon_\pi$ leads to an increasing uncertainty of $\langle \alpha_\mathrm{T} \rangle$ with rising $r_{\mathrm{K}/\pi}$. This simulation was used as well to calculate $\langle E_\mathrm{thr} \cos \theta \rangle = \unit[(1.34 \pm 0.18)]{TeV}$ for the location of the LNGS. Also this value agrees with the result of $\langle E_\mathrm{thr} \cos \theta \rangle = \unit[(1.30\pm0.16 )]{TeV}$ we obtained using the MUSIC/MUSUN simulation inputs.\\
Figure \ref{fig:RKPI} shows the experimental and theoretical values of $\alpha_\mathrm{T}$ as functions of $r_{\mathrm{K}/\pi}$. The experimental value of $\alpha_\mathrm{T}$ has a weak dependence on $r_{\mathrm{K}/\pi}$ since it enters into the calculation of the effective temperature $T_\mathrm{eff}$. To investigate this dependence, we calculated the daily $T_\mathrm{eff}$ for the same range of $r_{\mathrm{K}/\pi}$ values as above and redetermined $\alpha_\mathrm{T}$ for each set of $T_\mathrm{eff}$ values via the correlation to the measured muon flux as in section \ref{sec:Corr}. The resulting dependence is very weak and strongly overpowered by the statistical uncertainties of the measurements. 
\noindent
Finally, to determine the kaon-to-pion production ratio, we estimate the intersection of the two allowed $\alpha_\mathrm{T}$ bands to obtain a value of $r_{\mathrm{K/\pi}}=0.11^{+0.11}_{-0.07}$. The allowed region in $r_{\mathrm{K}/\pi}$ and $\alpha_\mathrm{T}$ has been determined by adding the $\chi^2$ profiles of the Borexino measurement and the theoretical prediction.\\
Former indirect measurements of the kaon-to-pion ratio were presented by the MINOS~\cite{MINOS} and IceCube~\cite{IceCube} experiments using a similar approach. Direct measurements have been carried out at accelerators, e.g. by STAR for Au+Au collisions at RHIC~\cite{AdlerAuAuRKPI}, by NA49 for Pb+Pb collisions at SPS~\cite{AfanasievPbPbRKPI}, and by E735 for $\mathrm{p+\bar{p}}$ collisions at Tevatron~\cite{ppbarcoll}. Results of many older measurements using various reactions are summarized and referred to in~\cite{GaColl}. The theoretical uncertainty of the kaon-to-pion ratio in current cosmic ray models is of the order of $40\%$~\cite{RKPILit}. Even though the indirect measurements do not directly compare with the accelerator experiments since the latter are performed with fixed beam energies, the central values are consistent as shown in figure \ref{fig:comp_rkpi}. 
\noindent
We place the Borexino data point in figure \ref{fig:comp_rkpi} at a center-of-mass energy $\sqrt{s}=\unit[(190\pm 28)]{GeV}$, calculated assuming an average collision of a primary $\unit[18]{TeV}$ proton on a fixed nucleon target. The proton energy is chosen to be ten times the mean threshold energy $\langle E_\mathrm{thr}\rangle = \unit[(1.8\pm0.2)]{TeV}$ we computed using the MUSIC/MUSUN inputs in our simulation, given that cosmic muons with $E>\unit[1]{TeV}$ obtain on average one tenth of the energy of the primary cosmic ray particle~\cite{Gaisser}. Due to the broad energy range of contributing muons, uncertainties on the center-of-mass energy need to be considered for the indirect measurements. Our result agrees with former indirect and direct measurements. Note that while the indirect measurements feature larger uncertainties than the accelerator experiments, they may infer the atmospheric kaon-to-pion ratio using cosmic ray data. Due to the smaller muon statistics at greater depths, our measurement uncertainty is larger than for the MINOS and IceCube results. However, Borexino contributes the data point at the highest center-of-mass energy for indirect as well as fixed target measurements.
\section{Lomb-Scargle Analysis of Muon Flux and Temperature}
\label{sec:LS}
\begin{figure}[tbp]
\begin{minipage}{0.5\textwidth}
\includegraphics[width = \textwidth]{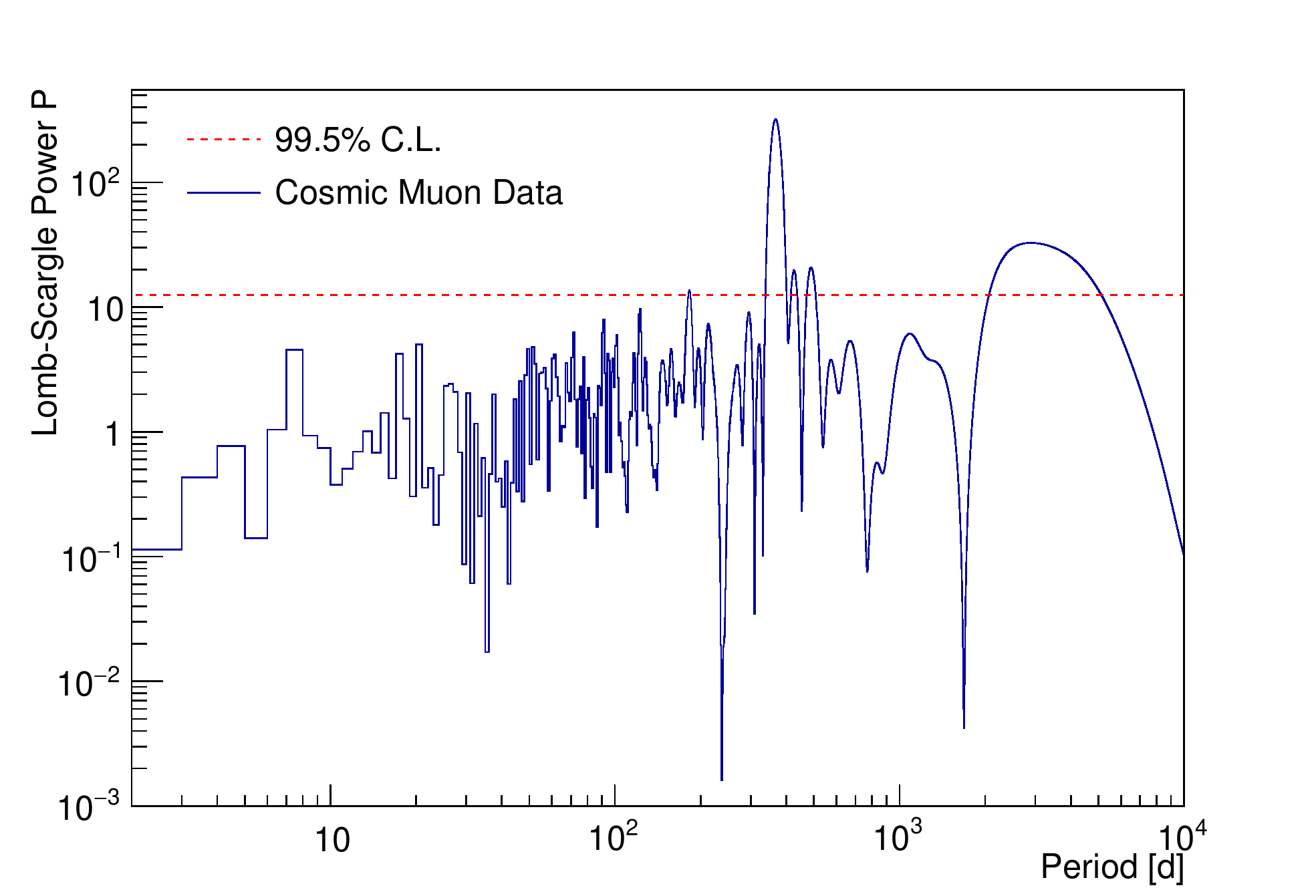}
\end{minipage}
\begin{minipage}{0.5\textwidth}
\includegraphics[width = \textwidth]{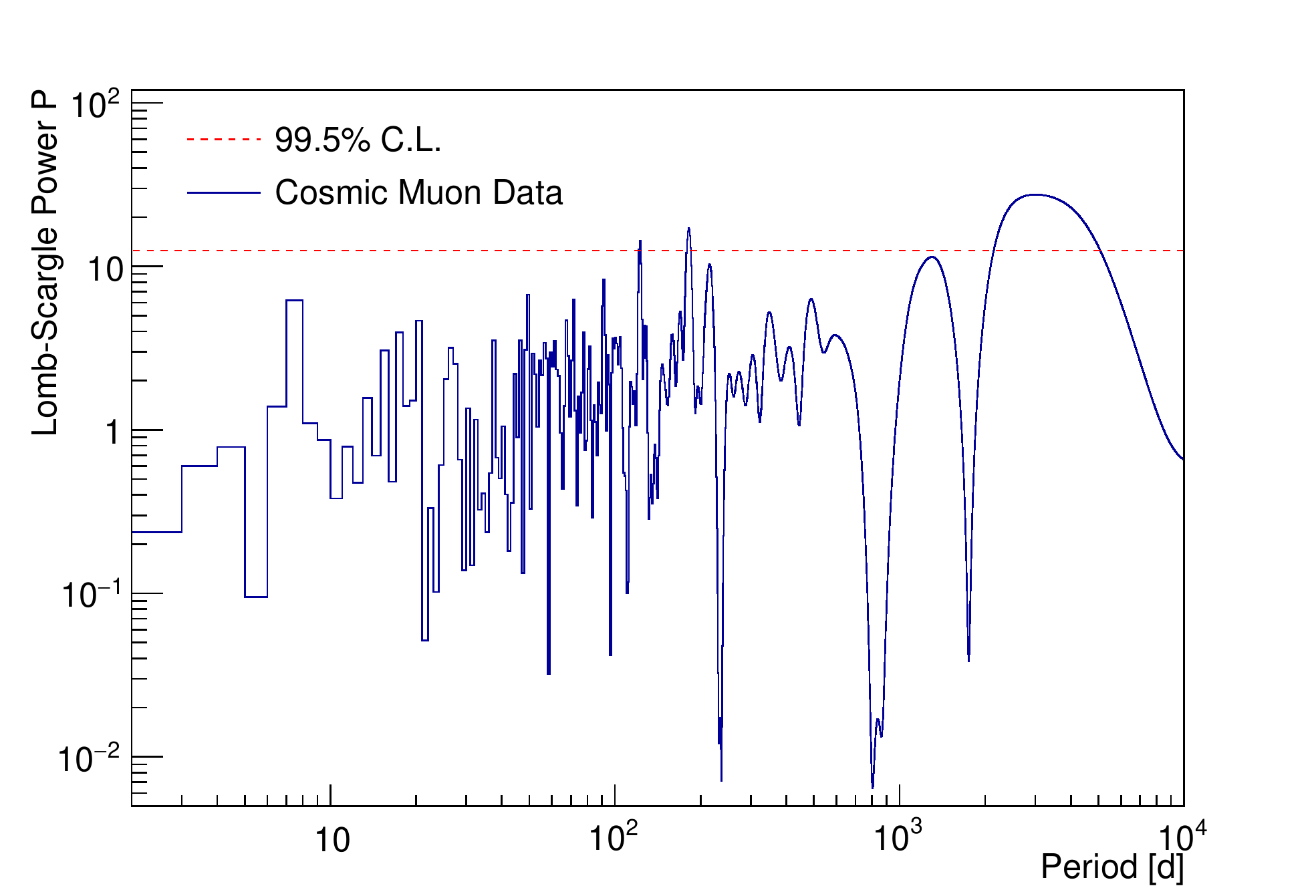}
\end{minipage}
\caption{\label{fig:muon_ls}The \textit{left} side shows the LS periodogram for the ten year cosmic muon data acquired with Borexino. The \textit{right} side shows the LS periodogram of the cosmic muon data after the seasonal modulation was subtracted statistically. The red lines indicate the significance level of $99.5\%$.}
\end{figure}

Besides the seasonal modulation of the cosmic muon flux underground, further physical processes might affect the cosmic muon flux and cause modulations of different periods. To investigate the presence of such non-seasonal modulations in the cosmic muon flux with Borexino, we perform a Lomb-Scargle analysis of the muon flux data.\\
Lomb-Scargle (LS) periodograms~\cite{Lomb,Scargle} constitute a common method to identify sinusoidal modulations in a binned data set described by 
\begin{equation}\label{eq:Oscillation}
N(t) = N_0 \cdot \left(1+A\cdot \sin \left(\frac{2\pi t}{T} +\phi\right)\right),
\end{equation}
where $N(t)$ is the expected event rate at time $t$ given the data set is modulated with a period $T$, a relative amplitude $A$, and a phase $\phi$. The LS power $P$ for a given period $T$ in a data set containing $n$ data points may be calculated via
\begin{equation}\label{eq:LS_eq}
P(T) = \frac{1}{2\sigma^2}  \left( \frac{\left[\sum_\mathrm{j}^n w_\mathrm{j} (N(t_\mathrm{j})-N_0)\cos\frac{2\pi}{T}(t_\mathrm{j}-\tau)\right]^2}{\sum_\mathrm{j}^n \cos^2 \frac{2\pi}{T}(t_\mathrm{j}-\tau)}+
\frac{\left[\sum_\mathrm{j}^n w_\mathrm{j} (N(t_\mathrm{j})-N_0)\sin\frac{2\pi}{T}(t_\mathrm{j}-\tau)\right]^2}{\sum_\mathrm{j}^n \sin^2 \frac{2\pi}{T}(t_\mathrm{j}-\tau)}  \right),
\end{equation} 
where $N(t_\mathrm{j})-N_0$ is the difference between the data value in the j$^\mathrm{th}$ bin and the weighted mean of the data set $N_0$ and $\sigma^2$ is the weighted variance. The weight $w_\mathrm{j} = \sigma_\mathrm{j}^{-2}/\langle \sigma_\mathrm{j}^{-2}\rangle$ of the j$^\mathrm{th}$ bin is computed as the inverse square of the statistical uncertainty of the bin divided by the average inverse square of the uncertainties of the data set. The phase $\tau$ satisfies~\cite{LS_SNO} 
\begin{equation}\label{eq:tau}
\tan\left(\frac{4 \pi}{T}\cdot \tau\right) = \frac{\sum_\mathrm{j}^n w_\mathrm{j} \sin(\frac{4\pi}{T}\cdot t_\mathrm{j})}{\sum_\mathrm{j}^nw_\mathrm{j} \cos(\frac{4\pi}{T}\cdot t_\mathrm{j})}.
\end{equation}
Since the quadratic sums of sine and cosine are used to determine the LS power, the latter is unaffected by the phase of a modulation as long as its period is short compared to the overall measurement time.\\
Figure \ref{fig:muon_ls} shows a LS periodogram for the ten year cosmic muon data acquired with Borexino. 
\noindent
To estimate the significance at which a peak in LS power exceeds statistical fluctuations, we use the known detector livetime distribution and mean muon rate to produce $10^4$ white noise spectra distributed equally to the data. We define a modulation of period $T$ to be significant if it surpasses a LS power $P_\mathrm{thr}$ that is higher than $99.5\%$ of the values found for white noise spectra. This threshold is indicated by the red line in figure \ref{fig:muon_ls}.\\
Besides the leading peak of the seasonal modulation at $\unit[365]{d}$, several secondary peaks are visible in figure \ref{fig:muon_ls}, the second most significant one being a long-term period of
$\sim \unit[3000]{d}$. However, it is known that  the LS method may identify harmonics of the leading modulation as significant~\cite{understandLS}. Therefore, only the highest significance peak can be safely regarded as physical. In order to clarify if the long-term modulation is physical, we subtract the seasonal modulation as in eq. \ref{eq:Seas} with the parameters returned by the fit described in sec. \ref{sec:Seas}. The power spectrum of the subtracted data set is shown in figure \ref{fig:muon_ls} (right). The peak at $\sim \unit[3000]{d}$ remains to be significant and now has the highest LS power, which verifies its presence in the data. Additionally, the peak at $\unit[180]{d}$ still exceeds the significance level, although only slightly. We consider also this peak to be of physical origin and related to the minor maxima in winter and spring described in section \ref{sec:Seas}, which determine a deviation from the purely sinusoidal behaviour as previously noted in~\cite{BxFlux}. Finally, a peak at the verge of significance is observed at $\sim \unit[120]{d}$.\\
With the period of the long-term modulation being close to our overall measurement time, the phase of the modulation is expected to affect the LS power. To investigate this, we artificially generated data samples including a seasonal and a long-term modulation of $\unit[3000]{d}$ period equally binned as the muon flux data. For each sample, the phase of the long-term modulation was altered and we computed a LS periodogram. We found the location of the peak to vary between $\sim 2550$\,d and $\sim 3750$\,d, which indicates the absolute uncertainty of the period. However, the long-term modulation appears as a significant peak in the LS periodogram independent of the inserted phase.\\
\begin{figure}[tbp]
\begin{minipage}{0.5\textwidth}
\includegraphics[width = \textwidth]{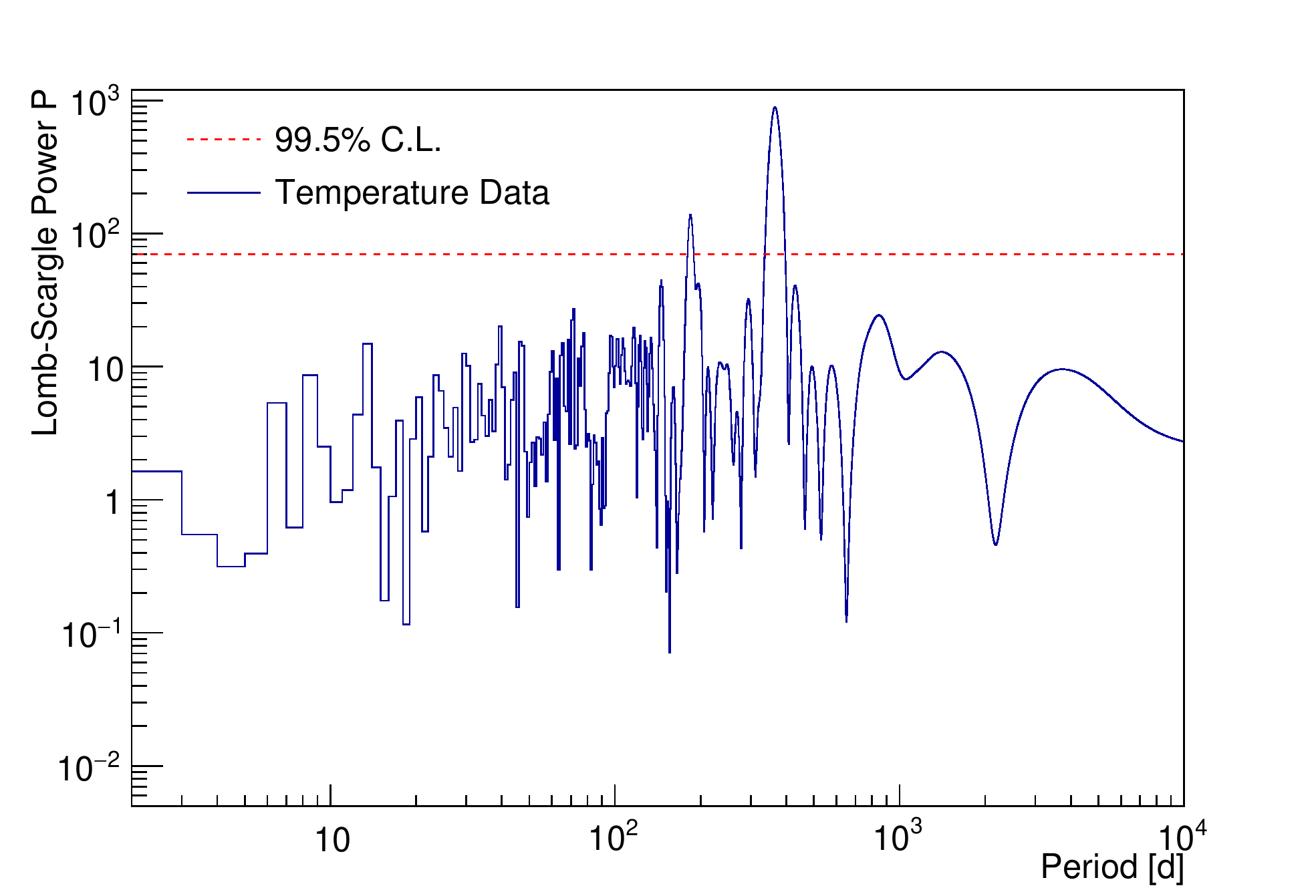}
\end{minipage}
\begin{minipage}{0.5\textwidth}
\includegraphics[width = \textwidth]{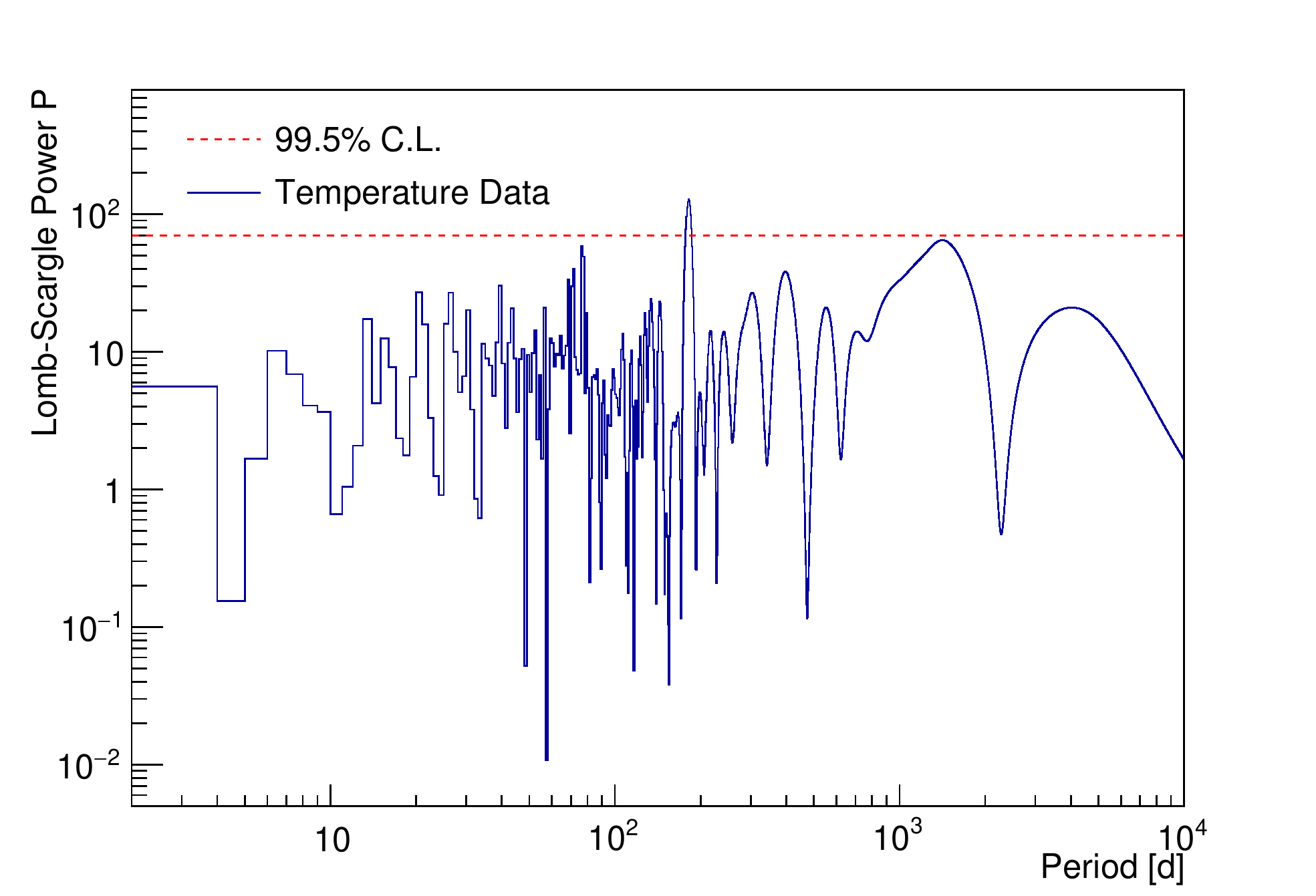}
\end{minipage}
\caption{\label{fig:ls_temp}The \textit{left} side shows the LS periodogram for the ten year effective atmospheric temperature data at the location of the LNGS~\cite{ecmwf}. On the \textit{right} side, the LS periodogram of the effective atmospheric temperature data after the seasonal modulation was subtracted statistically is shown. The red lines indicate the significance level of $99.5\%$.}
\end{figure}\noindent
On the left panel of figure \ref{fig:ls_temp}, we show the LS periodogram of the effective atmospheric temperature. Here, only the seasonal modulation and the $\unit[180]{d}$ period are found as significant peaks. No further period surpasses the threshold power.\\
To ensure that no long-term modulation might be inserted by the statistical subtraction of the seasonal modulation from the data, we repeated this procedure for the effective temperature data. As illustrated in the right panel of figure \ref{fig:ls_temp}, no further significant peaks are introduced by this approach. However, the $\unit[180]{d}$ period remains above the significance level, confirming our understanding of its origin. Thus, we conclude that the significant long-term modulation in the cosmic muon flux at $\sim \unit[3000]{d}$ is not present in and, hence, not related to the effective atmospheric temperature.\\
We determine the phase and amplitude of the observed long-term modulation by fitting a function accounting for both the seasonal and the long-term modulation of the form
\begin{equation}\label{eq:long_mod}
I_\mu(t) = I_\mu^0+\Delta I_\mu = I_\mu^0 +\delta I_\mu \cos\left( \frac{2\pi}{T}(t-t_0)\right) +\delta I_\mu^{\mathrm{long}} \cos\left( \frac{2\pi}{T^{\mathrm{long}}}(t-t_0^{\mathrm{long}})\right)
\end{equation}
to the daily-binned data. The fit returns a long-term modulation with a period $T^{\mathrm{long}}=\unit[(3010 \pm 299)]{d}=\unit[(8.25\pm0.82)]{yr}$, a phase $t_0^{\mathrm{long}}=\unit[(1993 \pm 271)]{d}$, and an amplitude $\delta I_\mu^{\mathrm{long}}=\unit[(14.7 \pm 1.8)]{d^{-1}}=(0.34 \pm 0.04)\%$. $T^{\mathrm{long}}$ is in good agreement with the period inferred from the LS periodogram and the phase of the long-term modulation indicates a maximum of the modulation in June 2012 for the investigated time frame. The parameters describing the seasonal modulation were left free in the fit and consistent results to the values reported in section \ref{sec:Seas} were obtained. The $\chi^2$/NDF reduces from 3921/3214, when only a single modulation according to eq. \ref{eq:Seas} is fitted to the data, to 3855/3211.\\
\begin{figure}[tbp]
\includegraphics[width = \textwidth]{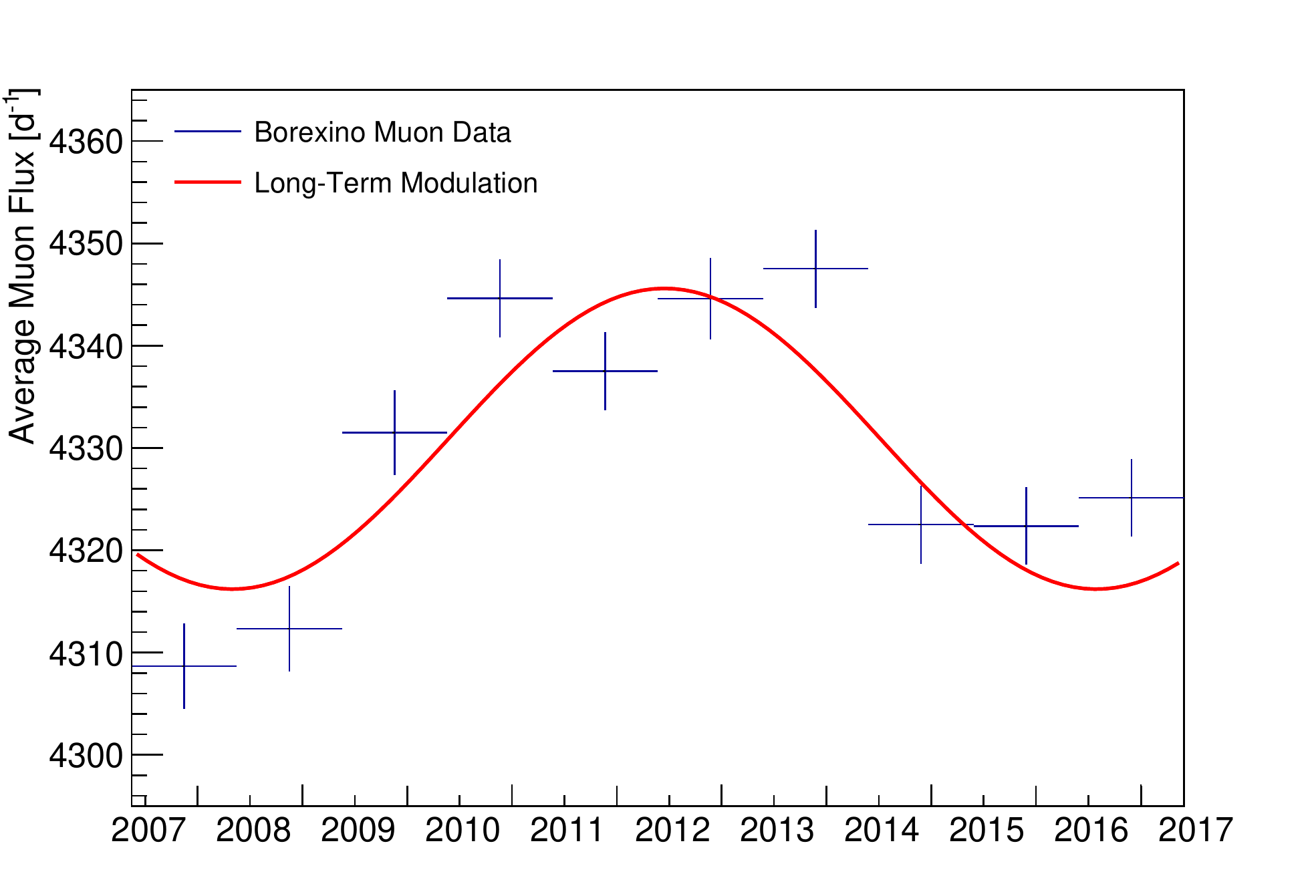}
\caption{\label{fig:ResFit}Cosmic muon flux measured by Borexino after statistically subtracting the leading order seasonal modulation in year-wide bins. The red line depicts the observed long-term modulation.}
\end{figure}\noindent
Figure \ref{fig:ResFit} shows our residual muon data in year-wide bins after having statistically subtracted the seasonal modulation in each day bin. Small effects of a possibly uneven distribution of the detector livetime across different years are thus removed. The red line shows the observed long-term modulation with the parameters as obtained by the fit to the daily-binned data. The data points show a clear variation in time, fully consistent with the fit result and the period observed in the LS analysis.

\section{Long-Term Modulation of the Cosmic Muon Flux and the Solar Activity}
\label{sec:SolarAct}
\begin{figure}[tbp]
\begin{minipage}{.49 \textwidth}
\includegraphics[width = \textwidth]{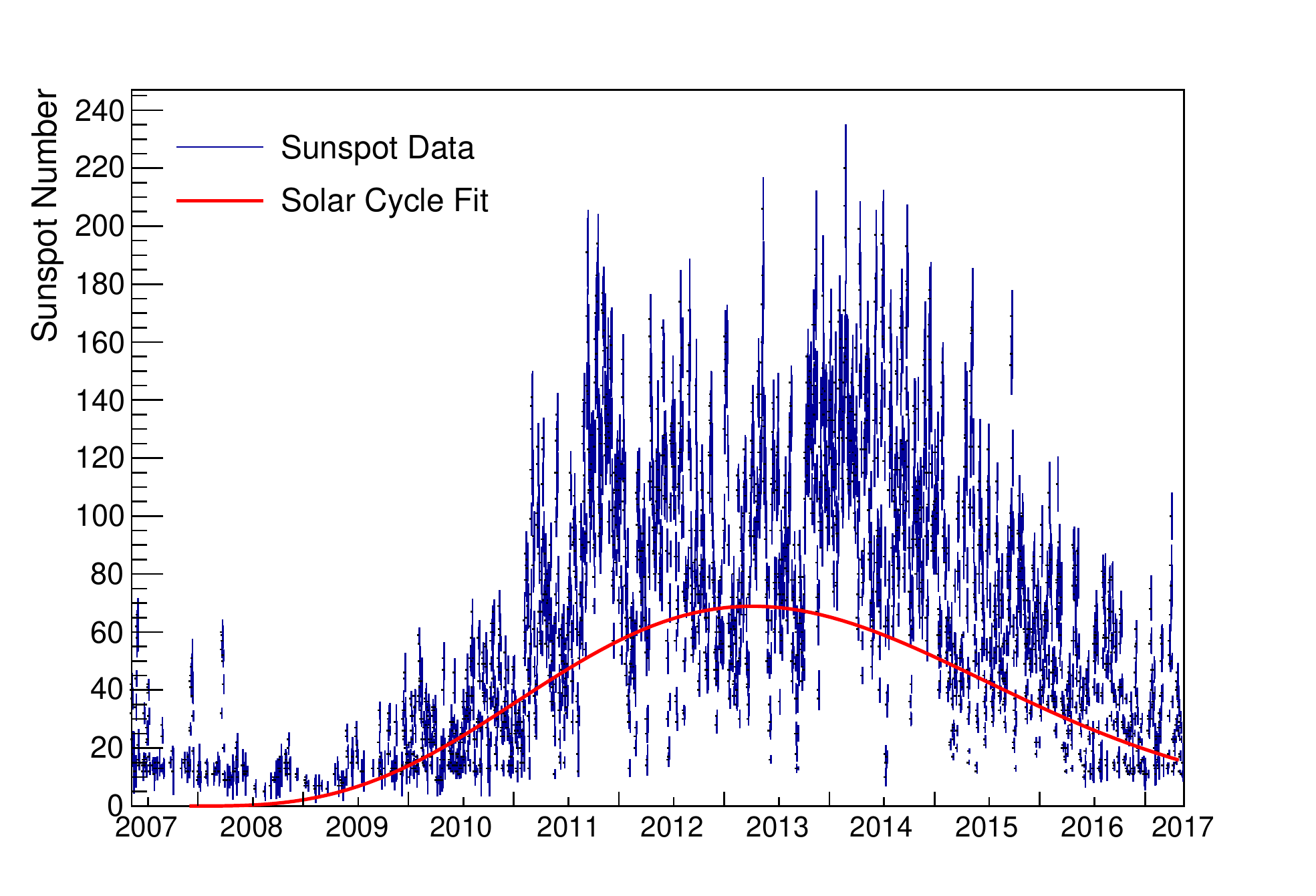}
\caption{\label{fig:sun_fit}Daily sunspot data corresponding to the Borexino data acquisition time~\cite{sunspot}. The curve shows a fit to the individual solar cycle.}
\end{minipage}\hspace{.02\textwidth}
\begin{minipage}{.49 \textwidth}
\vspace*{.4cm}
\includegraphics[width = \textwidth]{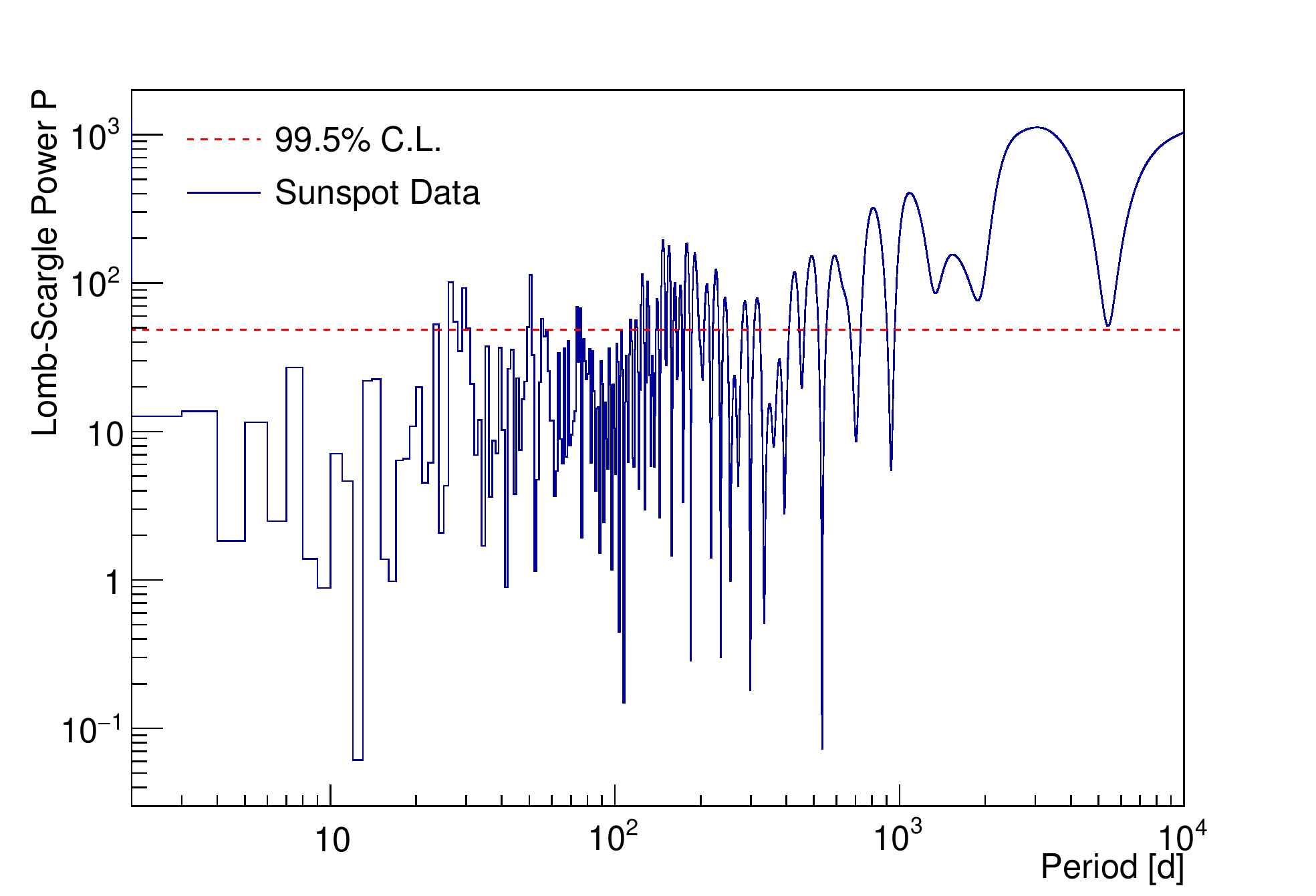}
\caption{\label{fig:ls_sun}LS periodogram of the daily sunspot data~\cite{sunspot} corresponding to the time frame of the Borexino muon data acquisition. The red line indicates the significance level of $99.5\%$.}
\end{minipage}
\end{figure}

A long-term modulation of the cosmic muon flux has been observed before, e.g. in ~\cite{LVD_long} and in~\cite{muon_puzzle}, also in comparison with the solar activity.
To investigate the possibility of such a correlation, we perform a LS analysis of the daily sunspot data provided by the World Data Center SILSO, the Royal Observatory of Belgium in Brussels~\cite{sunspot} for the timeframe corresponding to the cosmic muon data acquired by Borexino as shown in figure \ref{fig:sun_fit}. Since individual solar cycles are known to have significantly varied periods, it is not sensible to use a data set including earlier sunspot data. In figure \ref{fig:ls_sun}, the most significant peak in the LS periodogram occurs at a period of $\sim \unit[3000]{d}$, in coincidence with the long-term modulation of the cosmic muon flux. The significance level of $99.5\%$ was calculated following the procedure outlined in section \ref{sec:LS}. Further modulation periods are found to be significant in the sunspot data. This is expected since several authors observed minor modulations in the solar activity besides the solar cycle (see~\cite{SolarCycle} and refs. therein). Also the increase in LS powers towards very high periods is expected since solar activity modulations larger than the solar cycle have been observed.\\
\noindent
An individual solar cycle can be described by a cubic power law and a Gaussian decline as~\cite{SolarFit}
\begin{equation}\label{eq:solar_cycle_fit}
F(t)=A \left(\frac{t-t_\mathrm{s}}{b}\right)^3 \left[ \exp \left(\frac{t-t_\mathrm{s}}{b}\right)^2-c\right]^{-1}
\end{equation} 
with $A$ being the amplitude of the solar cycle, $t_\mathrm{s}$ the starting time, $b$ the rise time, and $c$ an asymmetry parameter. We fit the sunspot data starting from the minimum of solar activity in March 2009 accordingly to eq. \ref{eq:solar_cycle_fit} and obtain an amplitude $A=(168.1 \pm 0.4)$ sunspots per day, a start time $t_s = \unit[(-434 \pm 3)]{d}$ prior the minimum set at the 1st of March 2009, a rise time $b=\unit[(1578\pm 3)]{d}$, and an asymmetry parameter $c=0.000 \pm 0.001$. These parameters correspond to a maximum of the solar activity for the present cycle around the 8$^\mathrm{th}$ of April 2013. The $\chi^2/\mathrm{NDF}$ of the fit is 143145/2672, revealing that eq. \ref{eq:solar_cycle_fit} is only a first approximation of the complex sunspot data. Note that the comparably short period indicated by the fit matches optical observations of the current solar cycle~\cite{stce}.\\
When we apply a fit similar to eq. \ref{eq:solar_cycle_fit} to the measured cosmic muon flux after statistically subtracting the seasonal modulation, we obtain an amplitude $A=(45.4 \pm 8.1)$ muons per day, a start time $t_\mathrm{s} = \unit[(-34 \pm 111)]{d}$ prior the minimum set at the 1st of March 2009, a rise time $b=\unit[(1208\pm 116)]{d}$, and an asymmetry parameter $c=1.00 \pm 0.02$ at a $\chi^2/\mathrm{NDF}=3302/2665$. This indicates a maximum in March 2012.\\
In summary, we observe the following parameters for the long-term modulation of the cosmic muon flux and the solar sunspot activity:
\begin{center}
\begin{tabular}{c|cc}
& Half Period/ Rise Time  [d]& Maximum \\ \hline
Muon Flux (Sinusoidal Fit) & $1505 \pm 150$ & 16th of June 2012 $\pm$ 271\,d\\
Muon Flux (Gaussian Fit) & $1207 \pm 116$ & 4th of March 2012 $\pm$ 180\,d\\
Solar Sunspot Activity & & \\ (Gaussian Fit) & $1578 \pm 3$ & 8th of April 2013 $\pm $ 5\,d\\
\end{tabular}
\end{center}
The large uncertainties of the parameters of the long-term modulation of the cosmic muon flux for both fits as well as the large reduced $\chi^2$ of the fit to the sunspot data indicate that the results need to be treated with care. A correlation between the solar sunspot activity and the flux of high energy cosmic muons can neither be ruled out nor clearly be proven. However, we find indications encouraging further investigation of this phenomenon, especially considering the agreement between the modulation periods observed in the LS analysis. To eventually prove or negate a correlation, longer measurement times for the underground muon flux across several solar cycles will be necessary.\\
Concerning the observation of a long-term muon flux modulation reported in ~\cite{muon_puzzle}, we note that: (1)  the amplitude of $(0.40\pm0.04)$\,\% is compatible with our observation; (2) the period is also in agreement with the duration of the solar cycle; (3) the phase is however anti-correlated with the sunspot data. While the analysis of ~\cite{muon_puzzle} includes not only MACRO and LVD but also Borexino data from \cite{BxFlux}, the latter contributes only to the last four years and seems to be in tension with the presented modulation fit.\\
An indication of a positive correlation between solar activity and the flux of high energy cosmic rays was found by the Tibet AS array~\cite{TibetAirShower} in observations of the size of the shadow the Sun casts on $\sim \unit[10]{TeV}$ cosmic rays. This shadow was observed to be modulated depending on the solar activity with the shadow shrinking by $\sim 50 \%$ at the maximum activity. Monte Carlo simulations performed by this collaboration for several solar surface models that predict the coronal magnetic field gave consistent results. However, the amplitude $\delta I_\mu^{\mathrm{long}}=(0.34 \pm 0.04)\%$ we measure for the long-term modulation of the cosmic muon flux is too high to leave a modulation of the solar shade as the sole explanation of a possible correlation.

\section{Modulation of the Cosmogenic Neutron Production Rate}
\label{sec:Neutron}
\begin{figure}[tbp]
\begin{minipage}{.49 \textwidth}
\includegraphics[width = \textwidth]{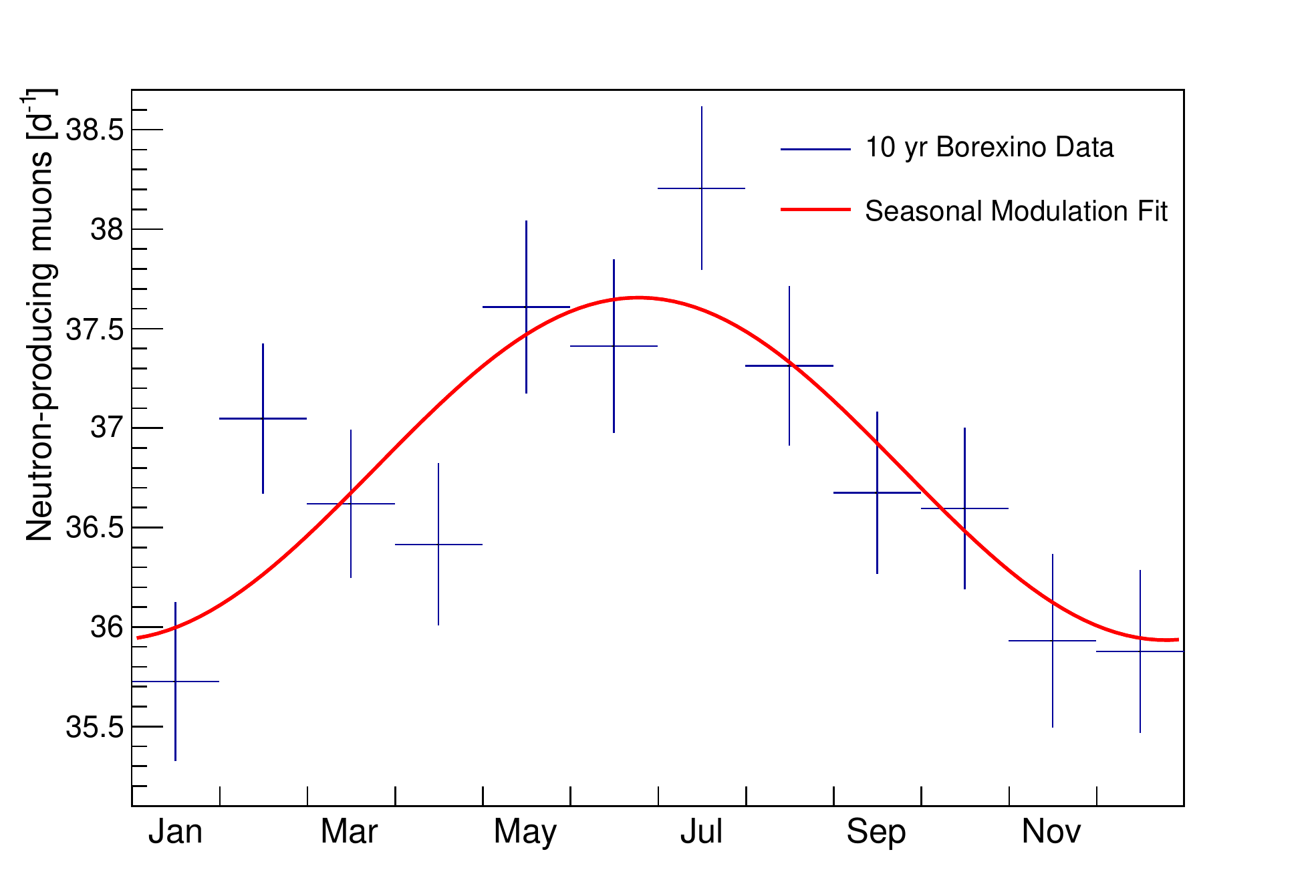}
\end{minipage}
\begin{minipage}{.49 \textwidth}
\includegraphics[width = \textwidth]{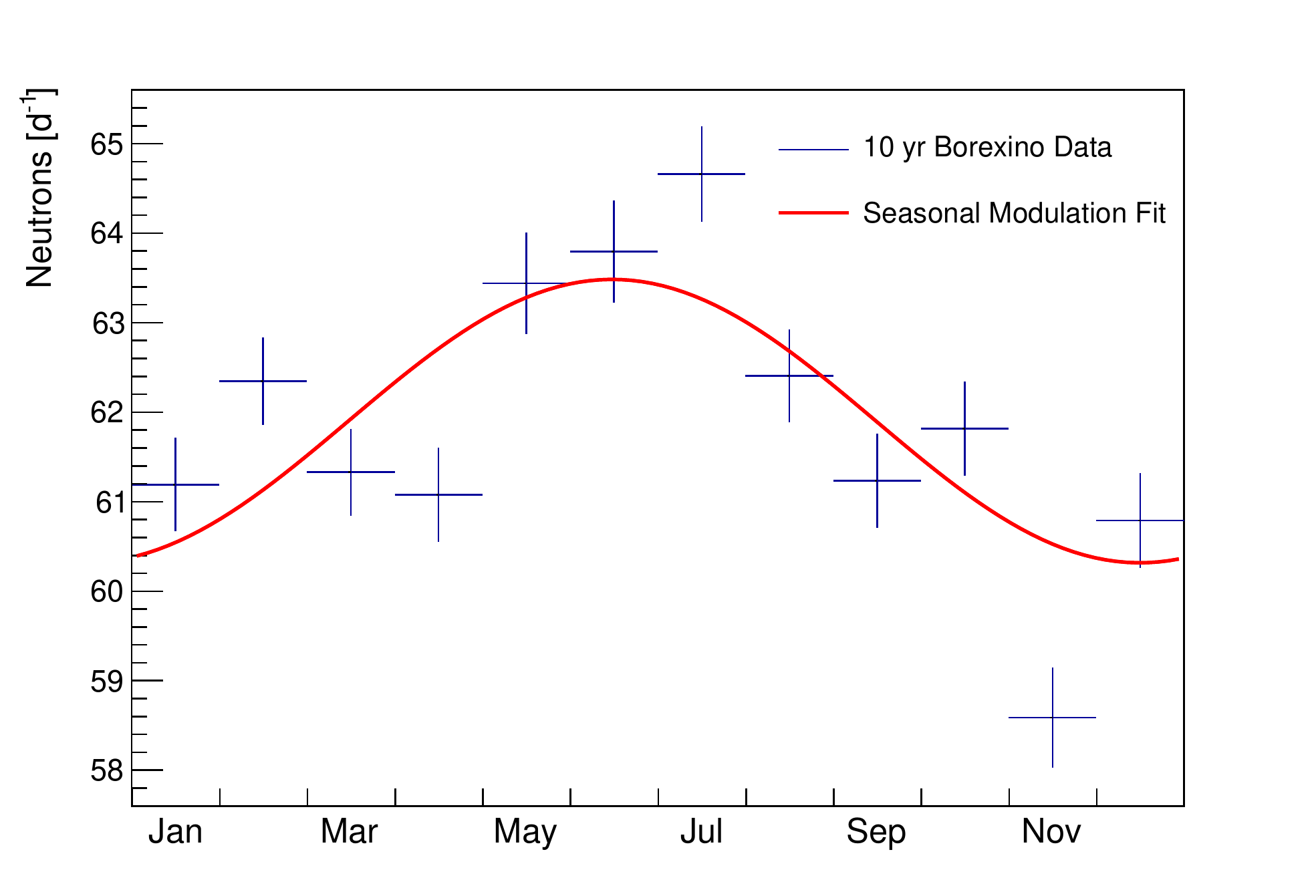}
\end{minipage}
\caption{\label{fig:neutron_mod}Rate of neutron-producing muons per day (\textit{left}) and the cosmogenic neutron production rate (\textit{right}) selecting only events with neutron multiplicity $n\leq 10$. Ten years of data are projected to a single year with monthly binning. The red line depicts a sinusoidal fit to the data with the period fixed to twelve months.}
\end{figure}

Cosmic muons may produce cosmogenic neutrons through various spallation processes on carbon in the Borexino organic scintillator target~\cite{Galbiati}. Neutrons are detected via the emission of a $\unit[2.2]{MeV}$ $\gamma$-ray following the capture on hydrogen or a total $\gamma$-ray energy of $\unit[4.9]{MeV}$ after the capture on $^{12}\mathrm{C}$. The capture time is $\tau = \unit[(259.7 \pm 1.3_\mathrm{stat} \pm 2.0_\mathrm{syst})]{\mu s}$~\cite{cosmogenics}. As a secondary product of cosmic muons, the number of cosmogenic neutrons is expected to also undergo a seasonal modulation. Cosmogenic neutrons have been discussed as a possible background for the expected modulation in direct searches for particle dark matter~\cite{vitalydama}. We investigate here the amplitude and phase of the cosmogenic neutron production rate.\\
We select cosmogenic neutrons in a special $\unit[1.6]{ms}$ acquisition gate that is opened after each ID muon~\cite{BxMuon}. Events with a visible energy corresponding to at least $\unit[800]{keV}$ are selected. The efficiency of the neutron selection has been measured to be $\epsilon_\mathrm{det} = (91.7 \pm 1.7_\mathrm{stat.} \pm 0.9_{ \mathrm{syst.}})\%$ after the stabilization of the electronics baselines $\sim \unit[30]{\mu s}$ after the passage of a muon. Due to the stable muon detection efficiency and no significant changes of the detector, we expect this efficiency to be stable.\\
We fail to observe the seasonal modulation of the cosmogenic neutron production rate, which we attribute to the occurrence of showering muons producing extremely high neutron multiplicities of up to $\sim 1000$~\cite{cosmogenics} and following a non-Poissonian probability distribution. This hypothesis is sustained by the fact that an annual modulation can indeed be seen in the LS periodogram for the rate of neutron-producing muons. However, in order for the modulation to be significant in the neutron production rate, we need to remove those neutrons that are produced in high multiplicity showers from the sample.\\
Figure \ref{fig:neutron_mod} shows on the left the monthly binned data of neutron-producing muons projected to one year with a sinusoidal fit similar to eq. \ref{eq:Seas} and the period fixed to twelve months.
Without efficiency correction, we obtain an average rate of neutron-producing muons $R_{\mu_\mathrm{n}}^0=\unit[(36.8 \pm 0.1)]{d^{-1}}$, an amplitude $\delta R_{\mu_\mathrm{n}}= \unit[(0.9 \pm 0.2)]{d^{-1}}=(2.3 \pm 0.5)\%$, and a phase $t_0 = \unit[(6.3 \pm 0.4)]{months}$. The reduced $\chi^2$ of the fit is $\chi^2/\mathrm{NDF}=11/9$. For the cosmogenic neutron production rate including neutrons produced in showers featuring up to 10 neutrons shown in figure \ref{fig:neutron_mod} on the right, we observe a rate $R_{\mathrm{n}}^0=\unit[(61.9 \pm 1.5)]{d^{-1}}$, an amplitude $\delta R_{\mathrm{n}}= \unit[(1.6 \pm 0.2)]{d^{-1}}=(2.6 \pm 0.4)\%$, and a phase $t_0 = \unit[(6.0 \pm 0.3)]{months}$ at a $\chi^2/\mathrm{NDF}=42/9$. Beyond a neutron multiplicity of 10, we are unable to observe the seasonal modulation applying the LS periodogram. The observed phase is in good agreement with the seasonal modulation of the entire cosmic muon flux. However, we find the amplitude of the modulation to be higher with a difference of about $2\sigma$ compared to the relative amplitude of $(1.36 \pm0.04)\%$ measured for the entire cosmic muon flux in section \ref{sec:Seas}.\\
Table \ref{tab:NeutronRes} lists the phase and amplitude of the seasonal modulation measured for the number of neutron-producing muons as well as for the neutron production rate applying increasingly high neutron multiplicity cuts. Consistent results are found for all samples with neutron multiplicities $n \leq 10$ beyond which the modulation is no longer significant in the LS periodogram. We find the phase of the cosmogenic neutron production rate to agree with the muon flux. Correspondingly, the maximum occurs approximately one month later than expected for dark matter particles. However, the amplitude of the modulation is higher compared to the muon flux's, independently of the multiplicity cut that was actually applied. To further probe this effect, we computed the average cosmogenic neutron production and neutron-producing muon rates in three summer months and three winter months for each data set. We inferred the amplitude by assuming a sinusoidal modulation around the mean of the two values. The amplitudes observed following this residual approach are listed in the last column of table \ref{tab:NeutronRes}. We find consistent values to the ones obtained from the fit in each data sample confirming the increased modulation amplitude.\\
\begin{table}
\centering
\begin{tabular}{|c|c|c|c|c|} \hline
Neutron &  Phase & Amplitude & Amplitude\\
Multiplicity & Projected $[\mathrm{months}]$ & Projected $[\%]$ & Residual $[\%]$\\ \hline
Neutron-producing muons & $6.3 \pm 0.4$ & $2.3 \pm 0.5$ & $2.5 \pm 0.8$ \\
$n = 1 $ & $6.6 \pm 0.5$ & $2.3 \pm 0.6$ & $2.6\pm 1.0$ \\
$n \leq 2$ & $6.1 \pm 0.3$ & $2.6 \pm 0.5$ & $2.8 \pm 0.8$ \\
$n \leq 3$ & $5.8 \pm 0.4$ & $2.2 \pm 0.4$ & $2.5\pm 0.8$ \\
$n \leq 4$ & $6.0 \pm 0.3$ & $2.2 \pm 0.4$ & $2.3 \pm 0.7$\\
$n \leq 5$ & $6.1 \pm 0.3$ & $2.3 \pm 0.4$ & $2.4 \pm 0.7$ \\
$n \leq 10$ & $6.0 \pm 0.3$ & $2.6 \pm 0.4$ & $2.7 \pm 0.7$\\ 
\hline
\end{tabular}
\caption{Parameters of the seasonal modulation observed for the number of neutron-producing muons and the neutron production rate applying increasingly high neutron multiplicity cuts. The second and third column show the phase and the amplitude of the seasonal modulation observed in the fit to the projected data, respectively. The last column shows the relative amplitude of the modulation following the residual approach.}\label{tab:NeutronRes}
\end{table}\noindent
The modulation of the cosmogenic neutron production rate has formerly been measured by the LVD experiment reporting an even higher modulation amplitude of $\delta R_\mathrm{n} = (7.7 \pm 0.8)\%$~\cite{LVDNeutrons}. Since the neutron production depends on the muon energy, the larger amplitude of the modulation of the cosmogenic neutron production rate was interpreted as an indirect measurement of a seasonal modulation of the mean energy $\overline{E}_\mu$ of cosmic muons observed in the LNGS. The values measured by LVD implied a modulation amplitude of the mean muon energy of $\sim 10\%$ or $\sim \unit[28]{GeV}$~\cite{LVDNeutrons}. 
Following this interpretation and the calculation outlined in~\cite{MeanEChange}, our measurement of the modulation amplitude of the cosmogenic neutron production rate would indicate a modulation of the mean energy of cosmic muons of $\sim \unit[4.5]{GeV}$. 
In order to verify the plausibility of this hypothesis, we have increased the mean muon energy by 1~GeV in the MUSIC/MUSUN~\cite{MUSUN} simulation codes by altering the muon spectral index (see eq.~\ref{eq:muonspec}) by 0.01. 
Even this small variation in the mean energy results in a 14\% change in the muon flux, which is ten times larger than the amplitude of the observed annual modulation. 
We conclude that a modulation of the mean muon energy of several GeV is unlikely.\\
Additionally, we performed simulations of the muon production using the MCEq code~\cite{MCEq} for various atmospheric models and inferred the muon surface spectra at Gran Sasso in winter and summer. We used these spectra as inputs for MUSIC/MUSUN to predict the corresponding underground spectra. Based on these simulations, we obtained a difference of the cosmic muon flux between the summer and winter spectra of about $1.4\%$ in accordance to our measurement but observed only slight deviations of the mean muon energy of less than $\sim \unit[0.1]{GeV}$.\\
With the exclusion of a modulation of the mean muon energy, a more complex energy dependence of the cross section for neutron production than the conventionally assumed $\overline{E}_\mu^{\alpha}$ law~\cite{MeanEChange} must be hypothesized to explain our observations. However, relatively large uncertainties of the atmospheric models and of the neutron production processes in the scintillator make it difficult to further investigate this percent-level effect.
\section{Conclusions}
\label{sec:Concl}
We have presented a new precision measurement of the cosmic muon flux in the LNGS under a rock coverage of $\unit[3800]{m\,w.e.}$ using ten years of Borexino data acquired between May 2007 and May 2017. We have measured a cosmic muon flux of $\unit[(3.432 \pm 0.001)\cdot 10^{-4}]{m^{-2}s^{-1}}$ with minimum systematics due to the spherical geometry of the detector. The seasonal modulation of the flux of high energy cosmic muons is confirmed and we have observed an amplitude of $(1.36\pm 0.04)\%$ and a phase of $\unit[(181.7 \pm 0.4)]{d}$ corresponding to a maximum on the 1$^\mathrm{st}$ of July. We have used data from global atmospheric models to investigate the correlation between variations of the muon flux and variations of the atmospheric temperature and showed that the seasonal modulation is also present in the effective atmospheric temperature. The correlation coefficient between the two data sets is 0.55 indicating a positive correlation and we have measured the effective temperature coefficient $\alpha_\mathrm{T} = 0.90 \pm 0.02$ reducing the statistical uncertainties of our former measurement by a factor $\sim 2$. The measurement is in good agreement with theoretical estimates and previous measurements carried out at the LNGS.\\ 
We have performed a Monte Carlo simulation to calculate the theoretical expectation of $\alpha_\mathrm{T}$ at the location of the LNGS as a function of the atmospheric kaon-to-pion ratio $r_{\mathrm{K}/\pi}$. By calculating the intersection region of the expected value and our measurement of $\alpha_\mathrm{T}$ in dependence on $r_{\mathrm{K}/\pi}$, we have indirectly measured $r_{\mathrm{K}/\pi} = 0.11^{+0.11}_{-0.07}$. This measurement is compatible with former indirect and accelerator measurements and constitutes a determination of $r_{\mathrm{K}/\pi}$ in a new energy region for fixed target experiments.\\
Based on a Lomb-Scargle periodogram, we have found evidence for a long-term modulation of the flux of high energy cosmic muons with a period of $\sim \unit[3000]{d}$ that is not present in the effective atmospheric temperature data. The amplitude of the long-term modulation is measured to be $(0.34 \pm 0.04)\%$ and the maximum occurs around June 2012. We have found indications of an agreement between this modulation and the solar activity. However, given our short observation time compared to the period of the long-term modulation, these indications are only modest and further investigation especially based on longer measurements is required. Additionally, the physical reason for a correlation between the high energy part of the cosmic muon flux and the solar activity remains unclear.\\
We have analyzed the production rate of cosmogenic neutrons in the Borexino detector as well as of the number of neutron-producing muons and found a seasonal modulation in phase with the cosmic muon flux but increased amplitudes of $\sim (2.6 \pm 0.4)\%$ and $\sim (2.3 \pm 0.5)\%$, respectively. We have shown that a strong modulation of the mean muon energy underground as an explanation of this phenomenon is disfavored by performing simulations of the muon surface spectrum in summer and winter using the MCEq software code~\cite{MCEq} and simulating the corresponding underground spectra using the MUSIC/MUSUN codes~\cite{MUSUN}.

\appendix
\section{Effective Temperature Weight Functions}\label{ap:A}
The weights assigned to temperature measurements at different atmospheric depths $X_\mathrm{n}$ in eq. \ref{eq:teff} to compute the effective atmospheric temperature are defined as~\cite{MINOS}
\begin{equation}
\begin{split}
 W_\mathrm{n}^\pi(X_\mathrm{n}) &\equiv \frac{A^1_\pi  e^{-X_\mathrm{n}/\Lambda_\pi} (1-X_\mathrm{n}/\Lambda_\pi^\prime)^2}{\gamma +(\gamma +1) B_\pi^1 K(X_\mathrm{n})(\langle E_\mathrm{thr} \cos \theta \rangle / \epsilon_\pi)^2}, \\
 W_\mathrm{n}^\mathrm{K}(X_\mathrm{n}) &\equiv \frac{A^1_\mathrm{K}  e^{-X_\mathrm{n}/\Lambda_\mathrm{K}} (1-X_\mathrm{n}/\Lambda_\mathrm{K}^\prime)^2}{\gamma +(\gamma +1) B_\mathrm{K}^1 K(X_\mathrm{n})(\langle E_\mathrm{thr} \cos \theta\rangle  / \epsilon_\pi)^2}
 \end{split}
 \end{equation}
with
\begin{equation} 
K(X_\mathrm{n}) \equiv \frac{(1-X_\mathrm{n}/\Lambda_\mathrm{M}^\prime)^2}{(1-e^{-X_\mathrm{n}/\Lambda_\mathrm{M}^\prime})\Lambda_\mathrm{M}^\prime}/X_\mathrm{n}. 
\end{equation}
The parameters $A^1_{\mathrm{K}/\pi}$ describe the relative contribution of kaons and pions, respectively, and include the amount of inclusive meson production, the masses of mesons and muons, and the muon spectral index $\gamma$. The input parameters are $A^1_\pi=1$ and $A^1_\mathrm{K}= 0.38 \cdot r_{\mathrm{K}/\pi}$, where $r_{\mathrm{K}/\pi}$ is the atmospheric kaon-to-pion production ratio. The parameter $B_{\mathrm{K},\pi}^1$ considers the relative atmospheric attenuation length of the mesons, $E_\mathrm{thr}$ is the threshold energy a muon needs to possess to penetrate the rock overburden and reach the LNGS, and $\theta$ is the zenith angle from which a muon is arriving. The attenuation lengths for primary cosmic rays, pions, and kaons are $\Lambda_\mathrm{N}$, $\Lambda_\pi$, and $\Lambda_\mathrm{K}$, respectively, and  $1/\Lambda_\mathrm{M}^\prime \equiv 1/\Lambda_\mathrm{N}-1/\Lambda_\mathrm{M}$. $\epsilon_\pi = \unit[(114\pm 3)]{GeV}$ and $\epsilon_\mathrm{K}= \unit[(851\pm 14)]{GeV}$ are the critical meson energies separating the interaction and the decay regimes. Since $E_\mathrm{thr}$ depends on the direction from which a muon arrives at the LNGS due to the shape of the rock overburden, the median of the product of the threshold energy and the cosine of the zenith angle $\langle E_\mathrm{thr} \cos \theta\rangle$ is used for the computation of $T_\mathrm{eff}$. Based on our Monte Carlo simulation, $\langle E_\mathrm{thr} \cos \theta\rangle = \unit[(1.34\pm 0.18)]{TeV}$ at the location of the LNGS.

\acknowledgments

The Borexino program is made possible by funding from INFN (Italy), NSF (USA), BMBF, DFG (OB 168/2-1, WU742/4-1, ZU 123/18-1), HGF, and MPG (Germany), RFBR (Grants 16-02-01026 A, 15-02-02117 A, 16-29-13014 ofim, 17-02-00305 A), RSF (Grant 17-02-01009) (Russia), and NCN (Grant No. UMO 2013/10/E/ST2/00180) (Poland). We acknowledge the generous hospitality and support of the Laboratori Nazionali del Gran Sasso (Italy).


\end{document}